\documentclass[journal,twoside,web]{ieeecolor}

\usepackage{T_AutomaticControl}
\usepackage{cite}
\usepackage{amsmath,amssymb,amsfonts}
\usepackage{algorithm,algorithmic}
\usepackage{cases}
\usepackage{graphicx}
\usepackage{textcomp}
\usepackage{dsfont}
\usepackage{enumerate}

\usepackage[caption=false,font=footnotesize]{subfig}

\usepackage{theorem}
\theorembodyfont{\rmfamily}

\newtheorem{prob}{Problem}
\newtheorem{lem}{Lemma}
\newtheorem{defin}{Definition}
\newtheorem{theorem}{Theorem}
\newtheorem{prop}{Proposition}
\newtheorem{assum}{Assumption}

\renewcommand{\QED}{\QEDopen}

\newcommand{\hs}{& \hspace{-3mm}}

\newcommand{\im}[1]{{\rm im}\, {#1}}
\newcommand{\diag}[1]{{\rm D}({#1})}
\DeclareMathOperator*{\argmin}{arg\,min}

\begin{document}
\title{Distributed Design of Glocal Controllers\\via Hierarchical Model Decomposition}

\author{Hampei~Sasahara,~\IEEEmembership{Member,~IEEE,}
        Takayuki~Ishizaki,~\IEEEmembership{Member,~IEEE,}
        Jun-ichi~Imura,~\IEEEmembership{Senior Member,~IEEE,}
        Henrik~Sandberg,~\IEEEmembership{Member,~IEEE,}
        and~Karl Henrik Johansson~\IEEEmembership{Fellow,~IEEE}
\thanks{H.~Sasahara, H.~Sandberg, and K.~H.~Johansson are with the Division of Decision and Control Systems, KTH Royal Institute of Technology, Stockholm, SE-100 44 Sweden e-mail: \{hampei, hsan, kallej\}@kth.se.}
\thanks{T.~Ishizaki and J.~Imura are with the Graduate School
of Engineering, Tokyo Institute of Technology,
Tokyo, 152-8552 Japan e-mail: \{ishizaki, imura\}@sc.e.titech.ac.jp.}
\thanks{This work was supported by JST MIRAI Grant Number 18077648, Japan, JSPS KAKENHI, Japan Grant Number 18K13774, the Swedish Research Council (grant 2016-00861), the Swedish Foundation for Strategic Research
(project CLAS), the Swedish Civil Contingencies Agency (project
CERCES).}
\thanks{Manuscript received Xxx xx, 20xx; revised Xxx xx, 20xx.}}

\maketitle

\begin{abstract}
This paper proposes a distributed design method of controllers having a glocal (global/local) information structure for large-scale network systems.
Distributed design, independent design of all subcontrollers that constitute a structured controller, facilitates scalable controller synthesis.
While existing distributed design methods confine attention to the decentralized or distributed information structures,
this study addresses distributed design of glocal-structured controllers.
Glocal control exploits the nature that network system's behavior can typically be represented as a superposition of spatially local fluctuations and global interarea oscillations by incorporating a global coordinating subcontroller with local decentralized subcontrollers.
The key idea to distributed design of glocal controllers is to represent the original network system as a hierarchical cascaded system composed of reduced-order models representing the global and local dynamics, referred to as hierarchical model decomposition.
Distributed design is achieved by independently designing and implementing subcontrollers for the reduced-order models while preserving the cascade structure.
This paper provides a condition for existence of the hierarchical model decomposition, a specific representation of the hierarchical system, a clustering method appropriate for the proposed approach, and a robust extension.
Numerical examples of a power grid evidence the practical relevance of the proposed method.
\end{abstract}


\section{Introduction}

\IEEEPARstart{T}{he} recent development of cyber and physical technologies facilitates large-scale dynamical systems, but increases also the complexity of the network systems to be controlled~\cite{Rajkumar2012A,Ilic2010Modeling}.
For large systems, instead of placing a centralized controller operating the entire system, it is preferred to deploy subcontrollers each of which monitors and actuates a small network unit while communicating with the others and thereby reducing the control structure complexity.
A representative information structure is the decentralized structure, where each subcontroller measures output signals within the assigned local network and transmits its control signals to local actuators.
Another common structure is the distributed structure, which allows communication among the subcontrollers with their neighborhood for accomplishing more complicated tasks through coordination and cooperation.
By virtue of their sparse communication topology, implementation of such structured controllers is scalable, which leads to their broad applicability to large-scale systems~\cite{sandell1978survey,bakule2008decentralized,siljak2011decentralized}.

Nevertheless, for most conventional methods in the literature, design of such structured controllers is \emph{not} necessarily scalable owing to its implicit philosophy of \emph{centralized design}, where a unique authority designs the entire controller for a fixed network system.
In practical large-scale systems, there are often multiple subcontroller designers,
each of whom designs and implements a subcontroller according to her control policy independently of the others.
For example, a power grid is governed by multiple companies each of whom is responsible for managing a subgrid.
Accordingly, each controller for frequency regulation is independently designed and operated by each company~\cite{Larsen1981Applying}.
In consequence, even if a controller is optimally designed for the grid at some time instant, the control performance is no longer guaranteed once one of the companies changes its control policy.

As the opposite of centralized design, the notion of \emph{distributed design}, where each subcontroller is designed independently of the others, has been introduced~\cite{Langbort10}.
Despite its practical importance, few studies on distributed design are available in the literature owing to the technical difficulty that each subcontroller must be designed to be capable of handling variations of the other subcontrollers.
To overcome this obstacle, several sophisticated distributed design methods have been proposed over the last decade.
As a distributed design method of decentralized controllers, retrofit control has been proposed~\cite{Ishizaki2018Retrofit,Ishizaki2019Modularity,Sasahara2020Parameterization}.
Distributed design methods of distributed controllers having a general communication topology have also been proposed\cite{Langbort10,Farokhi2013Optimal,JIANG1996A,Dashkovskiy2010Small,Riverso2016Plug,Pates2017Scalable,
Qu2014Modularized,Wang2018Separable,Anderson2019System}.

With the aforementioned background, this study addresses the distributed design problem of controllers having a specific \emph{glocal (global/local)} structure.
Glocal control, originally proposed in~\cite{Hara15}, employs a structured controller inspired by the fact that network system's behaviors can typically be represented as a superposition of spatially global and local behaviors.
For instance, the behavior of a power grid can be decomposed into global interarea oscillations and local fluctuations~\cite{Chow85Time}.
To utilize this nature, a global coordinating subcontroller is combined with local decentralized subcontrollers in the glocal control framework.

The objective of this paper is to develop a distributed design method for glocal controllers.
To this end, we introduce \emph{hierarchical model decomposition}, a hierarchical cascaded representation whose upstream and downstream parts stand for local and global reduced-order models, respectively.
Hierarchical model decomposition is an alternative representation of the original network system to be controlled.
Our fundamental idea is to design and implement subcontrollers for the reduced-order models while preserving the cascade structure of hierarchical model decomposition.
Owing to the cascade structure, the stability of the entire closed-loop system can be guaranteed as long as every reduced-order model is stabilized by its corresponding subcontroller.
In this paper, we resolve technical issues related to this idea.

The main contributions of this paper are outlined as follows:
First, we propose a systematic method of distributed design of glocal controllers based on hierarchical model decomposition.
Specifically, we provide a necessary and sufficient geometric condition on the existence of a hierarchical model decomposition.
We further derive an implicit representation of all hierarchical model decompositions through linear matrix equations and show how the designed control policy can be implemented based on a functional observer.
Second, we develop a clustering algorithm that produces clusters appropriate for the proposed method based on a greedy approach.
Third, we extend the framework to the case where no exact hierarchical model decompositions exist.
To handle this situation, we introduce a robust hierarchical model decomposition with approximation error, where also the error dynamics is decomposed into a hierarchical form.
Fourth and finally, we illustrate the potential impact of our theoretical findings through practical examples.
In particular, we design a glocal controller via the proposed approach for the 48-machine NPCC (Northeast Power Coordinating Council) system~\cite{Chow2013Power}, a model of the power grid in New York and neighboring areas.
Preliminary versions of this work have been presented in~\cite{sasahara2017glocal,Sasahara2019Hierarchical}, but they did not include the clustering algorithm, the robust extension, and the power grid example.

\subsection*{Related work}
A few studies on distributed design of structured controllers can be found.
Before the notion of distributed design in~\cite{Langbort10}, similar problems have been discussed in~\cite{JIANG1996A,Dashkovskiy2010Small}.
The underlying idea is to reduce the norm of the input-output maps relevant to interactions among the subsystems and to guarantee the overall stability based on the small-gain theorem.
Distributed design of distributed controllers guaranteeing bounded-input bounded-output stability has been proposed in~\cite{Riverso2016Plug} on the premise that interaction signals are bounded.
Different from the small-gain approaches, retrofit control has been proposed for distributed design of decentralized controllers~\cite{Ishizaki2018Retrofit,Ishizaki2019Modularity,Sasahara2020Parameterization}.
Regarding distributed information structure, deadbeat control~\cite{Langbort10,Farokhi2013Optimal}, integral quadratic constraint approach~\cite{Pates2017Scalable}, passivity-based approach~\cite{Qu2014Modularized}, and system-level synthesis~\cite{Wang2018Separable,Anderson2019System} approaches have been proposed.

Glocal control has been introduced in~\cite{Hara15} based on~\cite{Hara2009Consensus,Tsubakino2012Eigenvector}.
One of its features is hierarchical structure with spatial multiple-resolutions.
The idea of glocal control is to implement multi-resolved subcontrollers.
Especially in power system control, considerable efforts have been devoted to designing controllers having hierarchical structure~\cite{Kamwa2001Wide}.
In the classical approach, referred to as multi-level control~\cite{Brucoli1984Decentralized,Rubaai1991Transient,Huang1991Two}, the control signal is decomposed for each machine into two components generated by subcontrollers at global and local levels.
More recently, wide-area control~\cite{Chakrabortty2019Wide,Aranya13} has attracted attention along with the advancement of wide-area measurement system technology with sophisticated phasor measurement units.
Accordingly, several applications based on wide-area control have been proposed, e.g.,~\cite{Sondbakhsh2017A,Zenelis2018Wide,Xue2019Control,Thakallapelli2019Alternating}.
However, distributed design of the hierarchical structured controllers has not been discussed so far in the power systems literature.

Finally, we emphasize that hierarchical model decomposition, which provides an equivalent system representation composed of reduced-order models, cannot be obtained with standard model reduction techniques, such as the projection-based model reduction~\cite{Antoulas2005Approximation} or the singular perturbation method through coordinate transformation~\cite{Kokotovic1999Singular}.
Indeed, our idea relies on modification of the state space, which can make the dimension of the proposed representation larger than that of the original system, although the models used for designing each subcontroller are decomposed.

\subsection*{Organization and Notation}
In Sec.~\ref{sec:mot}, we present an illustration of our proposed method using a simple second-order network system example,
and subsequently, the problem is formulated.
Sec.~\ref{sec:hier}, where the proposed distributed design via hierarchical model decomposition is presented, gives the main technical results related to our approach.
Based on these findings, Sec.~\ref{sec:clust} develops a clustering algorithm that produces clusters appropriate for the proposed design framework.
In Sec.~\ref{sec:rob}, we propose a robust extension of the proposed method.
Sec.~\ref{sec:num} verifies the theoretical findings and demonstrates its practical effectiveness through numerical examples of power grids.
Finally, Sec.~\ref{sec:conc} draws conclusion.

We denote the set of real numbers by $\mathbb{R}$,
the $n$-dimensional identity matrix by $I_n$,
the $n\times m$ zero matrix by $0_{n\times m}$,
the $n$-dimensional all-ones vector by $\mathds{1}_n$,
the vector where $x_i$ for $i\in \mathcal{I}$ are concatenated vertically by ${\rm col}(x_i)_{i \in \mathcal{I}}$, and
the block diagonal matrix whose diagonal blocks are composed of matrices $M_i$ for $i\in\mathcal{I}$ by $\diag{M_i}_{i \in \mathcal{I}}$.
The subscript for the variables is omitted when the dimension is clear from the context.
Moreover, we denote the Kronecker product by $\otimes$,
the transpose and a pseudoinverse of a matrix $M$ by $M^{\sf T}$ and $M^\dagger$, respectively,
the direct sum and the sum space of linear subspaces $\mathcal{X}$ and $\mathcal{Y}$ by $\mathcal{X}\oplus\mathcal{Y}$ and $\mathcal{X}+\mathcal{Y}$, respectively,
the image space of a matrix $M$ by $\im{M}$,
the set $\{y=Mx:x\in\mathcal{X}\}$ for a matrix $M$ and a set $\mathcal{X}$ by $M\mathcal{X}$, and
the controllable subspace with respect to the pair $(A,B)$ by $\mathcal{R}(A,B)$.
Appendix contains the proofs.

\section{Problem Formulation}
\label{sec:mot}
\subsection{Motivating Example}
\label{subsec:ex}
\begin{figure}[t]
\centering
\includegraphics[width = .98\linewidth]{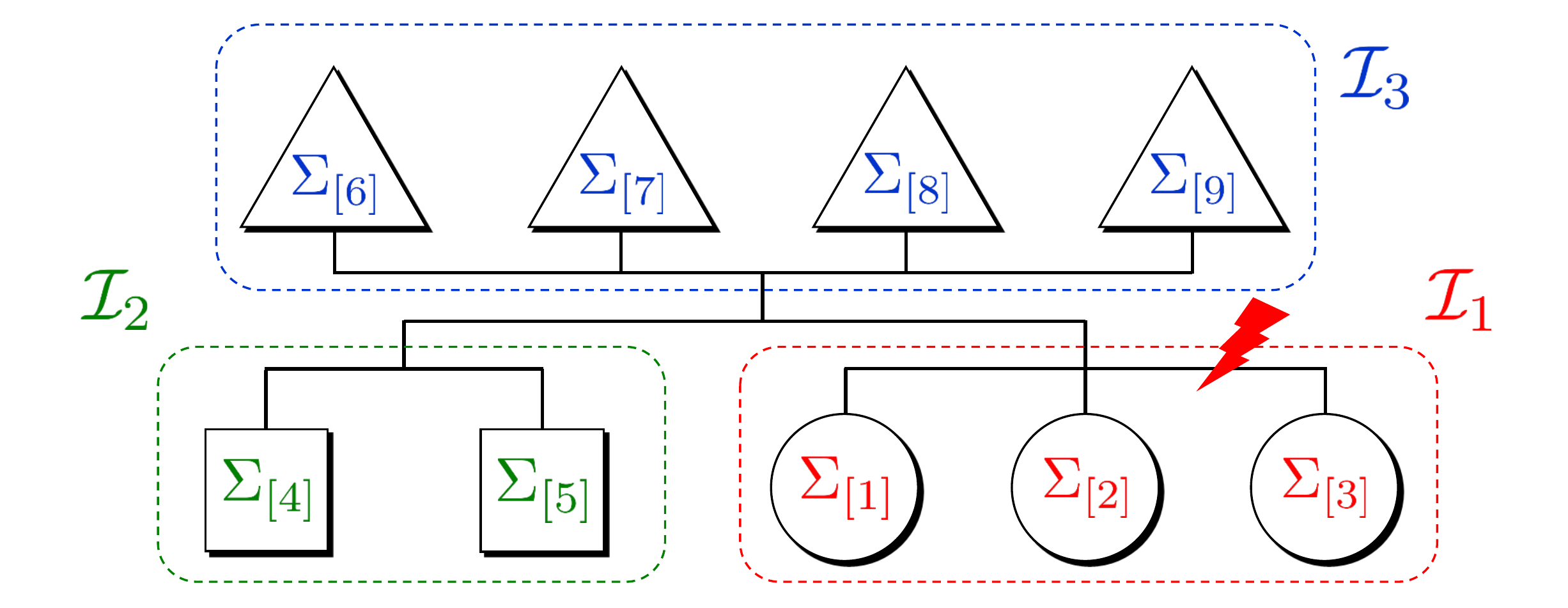}
\caption{Example: a network system where each component's dynamics is given as a second-order system.
The parameters of the components represented by the same shape are assumed to be identical.}
\label{fig:mot}
\end{figure}

This subsection provides an example that motivates us to introduce the glocal control structure to be presented.
We here consider the network system illustrated in Fig.~\ref{fig:mot} representing a power grid~\cite{Kundur94} where each component's dynamics is given as a second-order system.
For $k=1,\ldots,9$, each component $\Sigma_{[k]}$ is given by
\begin{equation}\label{eq:second}
\Sigma_{[k]}: m_{[k]}\ddot{\theta}_{[k]}+d_{[k]}\dot{\theta}_{[k]}+v_{[k]}+u_{[k]}=0,\quad y_{[k]}=\omega_{[k]}
\end{equation}
where $\theta_{[k]}\in \mathbb{R}$ and $\omega_{[k]}:=\dot{\theta}_{[k]}\in \mathbb{R}$ are the state,
$v_{[k]}\in\mathbb{R}$ is an interaction signal given by
\begin{equation}\label{eq:inter}
 \textstyle{v_{[k]} = \sum_{l \in \mathcal{N}_{[k]}} \alpha_{[k,l]}(\theta_{[k]}-\theta_{[l]}),}
\end{equation}
$u_{[k]} \in \mathbb{R}$ and $y_{[k]} \in \mathbb{R}$ are the control input and the measurement output, respectively,
and $\mathcal{N}_{[k]}$ represents the index set corresponding to the components connected to $\Sigma_{[k]}$.
Let the strength of the interaction among the components be given by $\alpha_{[k,l]}=1,$ for any $l,k =1,\ldots,9.$
Note that the underlying graph has a certain symmetry.
For instance, the components $\Sigma_1,\Sigma_2,$ and $\Sigma_3$ are equivalent from the viewpoint of the other components.

The glocal control framework addresses network systems whose components are partially synchronized.
Noticing the symmetry of the graph structure, we impose this property by assuming that symmetric components are partially homogeneous.
Specifically, the parameters of the components represented by the same shape in Fig.~\ref{fig:mot} are assumed to be identical and given as
\begin{equation}\label{eq:para_nominal}
 (m_{[k]},d_{[k]}) = \left\{
 \begin{array}{cl}
 (3,0.4), & k = 1,2,3,\\
 (2,0.3), & k = 4,5,\\
 (1,0.2), & k = 6,7,8,9.
 \end{array}
 \right.
\end{equation}

\begin{figure}[t]
\centering
\includegraphics[width = .98\linewidth]{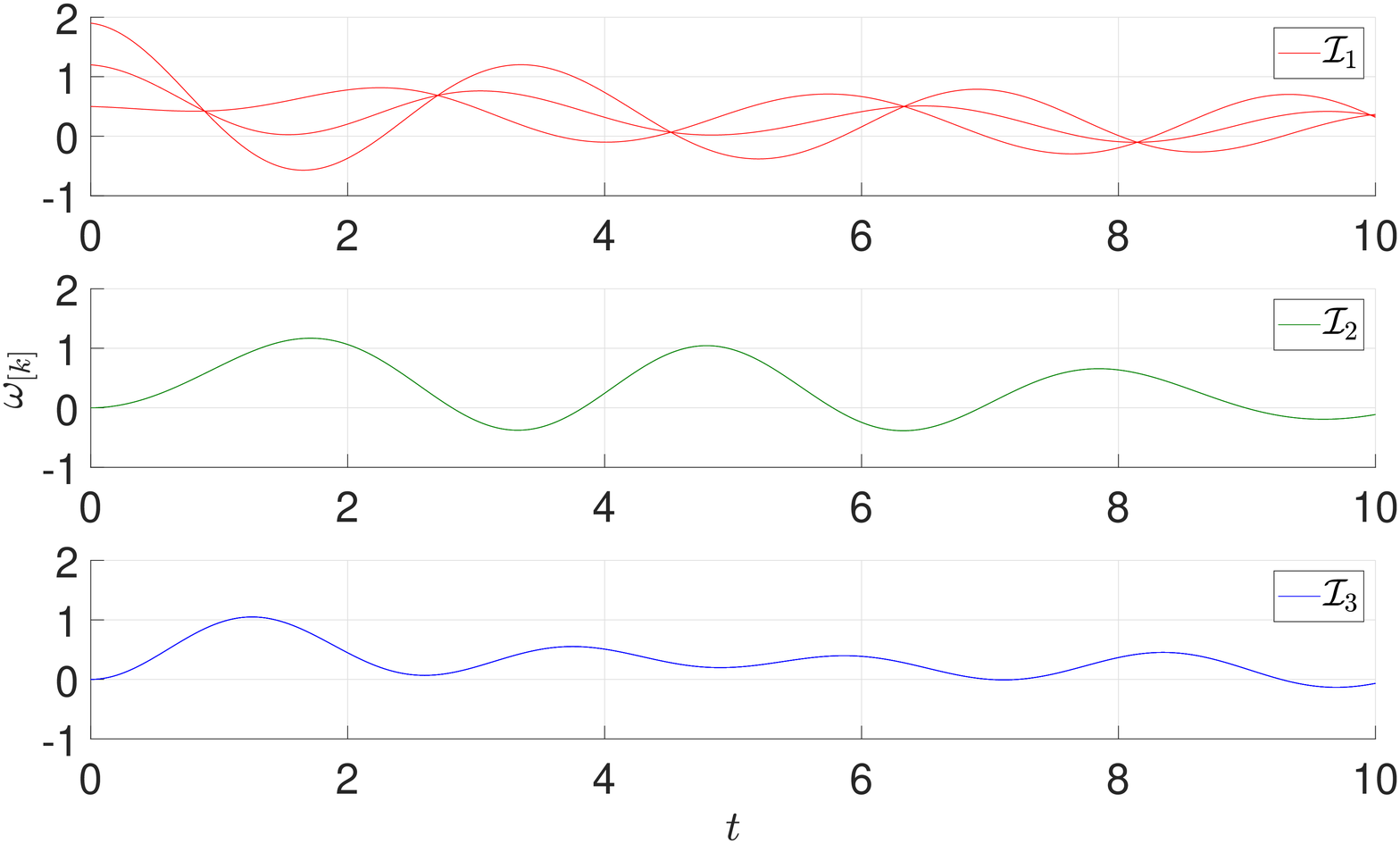}
\caption{Free response of the second-order system in Fig.~\ref{fig:mot} in response to an initial disturbance occurring inside $\mathcal{I}_1$.
The top, middle, and bottom correspond to $\mathcal{I}_1,$ $\mathcal{I}_2,$ and $\mathcal{I}_3$, respectively.
}
\label{fig:free}
\end{figure}

Then we can naturally arrange clusters as
\begin{equation}\label{eq:clu_ex}
 \mathcal{I}_1= \{1,2,3\},\quad \mathcal{I}_2= \{4,5\},\quad \mathcal{I}_3= \{6,7,8,9\},
\end{equation}
each of which contains homogeneous components.
Taking a single cluster,
we can expect that its components' responses to external signals injected into other clusters are the same owing to the partial homogeneity and network symmetry.
Fig.~\ref{fig:free} depicts the responses to a disturbance occurring inside $\mathcal{I}_1$ at the initial time.
As expected, the behaviors of $\Sigma_{[4]}$ and $\Sigma_{[5]}$, which belong to $\mathcal{I}_2$, are synchronized with each other.
Similarly, the components in $\mathcal{I}_3$ are synchronized.
This behavior can be interpreted as a global interarea oscillation among the clusters.
Thus, the entire state behavior in response to the local disturbance in $\mathcal{I}_1$ can be regarded as a superposition of global and local oscillations.
Further, because the system is linear, the same interpretation can apply for any disturbance by separating the disturbance to the sum of local disturbances.


This observation leads us to the glocal control structure.
Let the control input and the measurement output for the $i$th cluster be given by
\[
 u_i:={\rm col}(u_{[k]})_{k\in \mathcal{I}_i},\quad y_i:={\rm col}(y_{[k]})_{k\in \mathcal{I}_i},\quad i=1,2,3.
\]
Consider constructing the control input as a composition of global and local inputs where the global input maintains the states in each cluster synchronized.
Because the control inputs in each cluster affect the corresponding component's state in the same manner,
we take broadcast-type global control inputs as the global input.
Then the whole control input is given by
\[
\left[
 \begin{array}{c}
 u_1\\
 u_2\\
 u_3
 \end{array}
 \right]=
 \hat{\mathbf{u}}_{0} +
 \left[
 \begin{array}{c}
 \hat{u}_1\\
 \hat{u}_2\\
 \hat{u}_3
 \end{array}
 \right],\quad \hat{\mathbf{u}}_0 :=
 \left[
 \begin{array}{c}
 \mathds{1}_3 \hat{u}_{0,1}\\
 \mathds{1}_2 \hat{u}_{0,2}\\
 \mathds{1}_4 \hat{u}_{0,3}
 \end{array}
 \right].
\]
Similarly, the global measurement signal reads as $y_0:=[\mathds{1}_3^{\sf T} y_1\ \mathds{1}_2^{\sf T} y_2\ \mathds{1}_4^{\sf T} y_3]^{\sf T}$.
The global and local control inputs are basically used for suppressing global and local oscillations, respectively.
We design a glocal controller that contains global and local subcontrollers associated with these signals.

\subsection{System Description}
\label{subsec:sys}

Let us now introduce the general system description.
Consider a linear time-invariant interconnected system containing $N_0$ components
\begin{equation}\label{eq:component_k}
 \Sigma_{[k]}: \left\{
 \begin{array}{cl}
 \dot{x}_{[k]} \hs =A_{[k]}x_{[k]} + L_{[k]} \sum_{l \in \mathcal{N}_{[k]}} v_{[l]} + B_{[k]}u_{[k]}\\
 y_{[k]} \hs = C_{[k]}x_{[k]}
 \end{array}
 \right.
\end{equation}
for $k=1,\ldots,N_0$
with the interaction given by
\begin{equation}\label{eq:inter_components}
 \left[
 \begin{array}{c}
 v_{[1]}\\
 \vdots\\
 v_{[N_0]}
 \end{array}
 \right] = \left[
 \begin{array}{ccc}
 M_{[1,1]} & \cdots & M_{[1,N_0]}\\
 \vdots &  & \vdots\\
 M_{[N_0,1]} & \cdots & M_{[N_0,N_0]}
 \end{array}
 \right]
 \left[
 \begin{array}{c}
 x_{[1]}\\
 \vdots\\
 x_{[N_0]}
 \end{array}
 \right]
\end{equation}
with matrices $M_{[k,l]}$ for $k,l=1,\ldots,N_0$.
The signals $x_{[k]}, v_{[k]}, u_{[k]}, y_{[k]}$ denote the state, the interaction signal, the control input, and the measurement signal, respectively, and the set $\mathcal{N}_{[k]}$ denotes the neighborhood of $\Sigma_{[k]}$.
We assume that the dimensions of all the signals are identical with respect to $k$ and also that $u_{[k]}$ and $y_{[k]}$ are one-dimensional to avoid notational burden.

Next we introduce the glocal control structure for the given interconnected system.
First, for grouping the components, we arrange $N$ clusters $\mathcal{I}_i \subset \{1,\ldots,N_0\}$ for $i=1,\ldots,N$ that satisfy $\mathcal{I}_i\cap \mathcal{I}_j=\emptyset$ for $i\neq j$ and $\bigcup_{i=1}^{N} \mathcal{I}_i=\{1,\ldots,N_0\}$.
The clusters should be chosen such that the components in each cluster are synchronized in response to disturbance occurring outside.
The clustering problem will be addressed in Sec.~\ref{sec:clust}.

Given clusters, the subsystem regarding the $i$th cluster can be written as
\[
 \Sigma_i: \left\{
 \begin{array}{cl}
 \dot{x}_i \hs = A_ix_i + L_i \sum_{j \in \mathcal{N}_i} v_j + B_iu_i\\
 y_i \hs = C_i x_i,
 \end{array}\right.
\]
where the state is defined by $x_i:= {\rm col}(x_{[k]})_{k\in \mathcal{I}_i}$ and the other signals are defined in a similar manner, for $i=1,\ldots,N$.
The interaction among them is given by
\[
 \left[
 \begin{array}{c}
 v_1\\
 \vdots\\
 v_N
 \end{array}
 \right] = \underbrace{\left[
 \begin{array}{ccc}
 M_{1,1} & \cdots & M_{1,N}\\
 \vdots &  & \vdots\\
 M_{N,1} & \cdots & M_{N,N}
 \end{array}
 \right]}_{=:M}
 \left[
 \begin{array}{c}
 x_{1}\\
 \vdots\\
 x_{N}
 \end{array}
 \right].
\]
We denote the dimension of the state and the input in $\mathcal{I}_i$ by $n_i$ and $r_i$, respectively.


As in the motivating example, we let the control input be composed of global and local control inputs.
The following assumption is made.
\begin{assum}\label{assum:identical}
The input and output matrices in each cluster are identical, that is,
\[
 B_{[k]}=B_{[l]},\quad C_{[k]}=C_{[l]},\quad \forall k,l \in \mathcal{I}_i
\]
for $i=1,\ldots,N$.
\end{assum}
Accordingly, we form the control input as
\[
 {\rm col}(u_i)_{i=1}^N=\hat{\mathbf{u}}_0+{\rm col}(\hat{u}_i)_{i=1}^N,\quad \hat{\mathbf{u}}_0:=E_0\hat{u}_0
\]
with $E_0:=\diag{\mathds{1}_{r_i}}_{i=1}^N$ where $\hat{u}_0$ and $\hat{u}_i$ for $i=1,\ldots,N$ represent global and local control inputs, respectively.
Similarly, the global measurement signal is defined by $y_{0}:= E_0^{\sf T}y.$
Note that when Assumption~\ref{assum:identical} does not hold, it suffices to modify $E_0$ appropriately such that only the synchronized behavior in each cluster is excited by the modified global control input $E_0\hat{u}_0$.

The dynamics of \emph{the clustered interconnected system} is given by
\begin{equation}\label{eq:cl_sys}
 \left\{
 \begin{array}{cl}
 \dot{x} \hs = Ax+P_0B_0 \hat{u}_0+\sum_{i=1}^N P_i B_i \hat{u}_i\\
 y_0 \hs = C_0P_0^{\sf T} x\\
 y_i \hs = C_iP_i^{\sf T} x
 \end{array}
 \right.
\end{equation}
where $x := {\rm col}(x_i)_{i=1}^N$, $A:=\diag{A_i}_{i=1}^N+\diag{L_i}_{i=1}^NM$, and the matrices $P_0$ and $P_i$, which are consistent in broadcasting and embedding matrices, are defined by
\begin{equation}\label{eq:Ps}
\arraycolsep=1pt
 P_0 := \left[
 \begin{array}{ccc}
 \mathds{1}_{r_1}\otimes I_{n_{0,1}} & \cdots & 0_{n_1 \times n_{0,N}}\\
 \vdots & \ddots & \vdots\\
 0_{n_N\times n_{0,1}} & \cdots &\mathds{1}_{r_N} \otimes I_{n_{0,N}}
 \end{array}
 \right],\, P_i := \left[
 \begin{array}{c}
 0_{n_1 \times n_i}\\
 \vdots\\
 I_{n_i}\\
 \vdots\\
 0_{n_N \times n_i}
 \end{array}
 \right],
\end{equation}
and $n_{0,i}:=n_i/r_i$ is the dimension of the state of a subsystem in $\mathcal{I}_i$.
The matrices $B_0$ and $C_0$ are chosen such that
\[
 P_0B_0=\diag{B_i}E_0,\quad C_0P_0^{\sf T}=E_0^{\sf T}\diag{C_i}.
\]
Note that there always exist such $B_0$ and $C_0$ owing to Assumption~\ref{assum:identical}.
Finally, let $K_i$ be a subcontroller to be designed corresponding to $\hat{u}_i$ and $y_i$ for $i=0,1,\ldots,N$.
The specific information structure of the entire controller is discussed in Sec.~\ref{sec:hier}.

\subsection{Problem Formulation}

Based on the system description, we consider distributed design of a glocal controller, where each subcontroller can be designed independently of the others.
Let us introduce subcontroller sets $\mathcal{K}_i$ such that the entire closed-loop system is internally stable for any combination of subcontrollers in $\mathcal{K}_i$.
Distributed design of the subcontrollers is achieved by designing the $i$th subcontroller $K_i$ to be an element of $\mathcal{K}_i$ in the sense that $K_i$ can be chosen independently of the other subcontrollers.
\begin{prob}\label{prob:1}
Design a collection of subcontroller sets $\{\mathcal{K}_i\}_{i=0}^N$ such that the clustered interconnected system~\eqref{eq:cl_sys} with the subcontrollers $K_0,K_1,\ldots,K_N$ is internally stable for any choice of a tuple
\[
 (K_0,K_1,\ldots,K_N) \in \mathcal{K}_0\times\mathcal{K}_1\times\cdots\times\mathcal{K}_N.
\]
\end{prob}

In Problem~\ref{prob:1}, the subcontroller sets $\mathcal{K}_i$ are designed instead of subcontrollers $K_i$ themselves.
Note that a trivial solution can be given as singletons $\mathcal{K}_i=\{K_i\}$ where the unique controller stabilizes the entire system.
However, this choice is unfavorable for distributed design since the designed controller has no flexibility.
In this study, we seek for larger subcontroller sets.

\begin{figure*}[t]
\centering
\includegraphics[width = .98\linewidth]{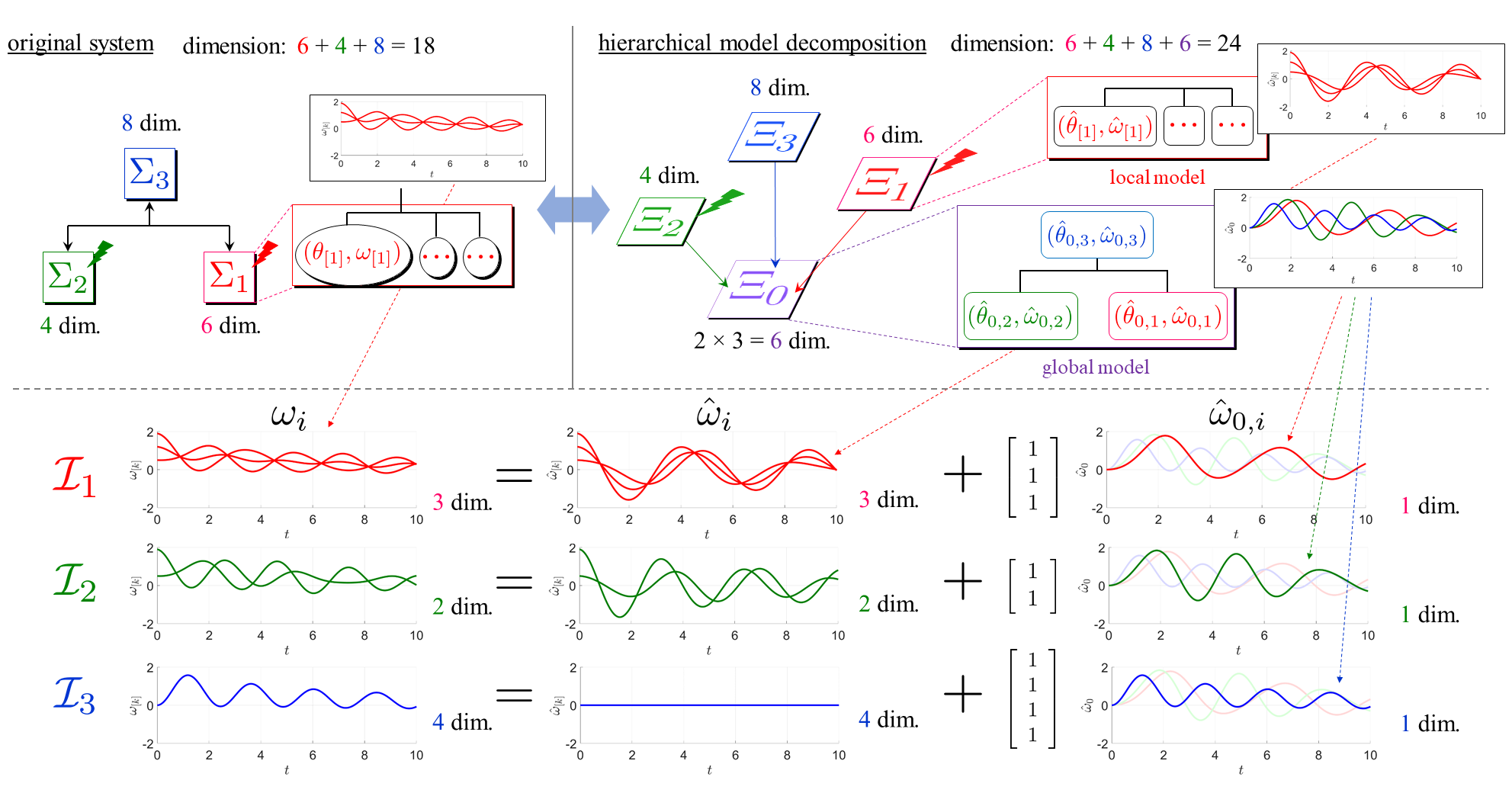}
\caption{
Hierarchical model decomposition of the second-order network system of the motivating example.
Top: block diagrams of the original network system and the hierarchical system.
Bottom: the responses of $\omega_i$ in $\Sigma_i$ and its representation as a superposition of $\hat{\omega}_i$ in $\Xi_i$ and $\hat{\omega}_{0,i}$ in $\Xi_0$ for $i=1,2,3$.
}
\label{fig:hier_sec_ex}
\end{figure*}

In Sec.~\ref{sec:hier}, we solve Problem~\ref{prob:1}, when the system and clusters are given in advance.
However, since the performance of the distributed design method depends on the choice of clusters, clustering should be included in the design process.
Thus we also address the following problem in Sec.~\ref{sec:clust} based on the results in Sec.~\ref{sec:hier}.
\begin{prob}\label{prob:2}
Develop a clustering method that produces proper clusters for the proposed distributed design method.
\end{prob}

\section{Distributed Design via Hierarchical Model Decomposition}
\label{sec:hier}


\subsection{Motivating Example Revisited}

This subsection describes our basic idea for distributed design of a glocal controller through the motivating example in Sec.~\ref{subsec:ex}.
As observed, any local disturbance occurring in a cluster excites only the synchronized behavior in the other clusters.
Thus behaviors of the network system in response to any disturbance can be interpreted as a superposition of global and local behaviors excited by the corresponding local disturbances.
The key idea is to derive a \emph{hierarchical} representation that explicitly describes these behaviors.


Consider describing the state variables as a superposition of global and local states, e.g.,
\begin{equation}\label{eq:super_ex1}
\underbrace{
\left[
\begin{array}{c}
 \omega_{[1]}\\
 \omega_{[2]}\\
 \omega_{[3]}
\end{array}
\right]}_{=:\omega_1} = 
\underbrace{
\left[
\begin{array}{c}
 \hat{\omega}_{[1]}\\
 \hat{\omega}_{[2]}\\
 \hat{\omega}_{[3]}
\end{array}
\right]}_{=:\hat{\omega}_1} + 
\left[
\begin{array}{c}
 1\\
 1\\
 1
\end{array}
\right]
\hat{\omega}_{0,1}
\end{equation}
where $\hat{\omega}_{[k]}$ for $k=1,2,3$ and $\hat{\omega}_{0,1}$ represent the local and global behaviors of $\omega_{[k]}$ in the first cluster, respectively.
Similarly, we consider $\hat{\omega}_i$ and $\hat{\omega}_{0,i}$ for the second and third clusters.
The global variable with respect to all clusters is denoted by $\hat{\omega}_0:={\rm col}(\hat{\omega}_{0,i})_{i=1,2,3}$.
The variables with respect to $\theta_i$, denoted by $\hat{\theta}_0$ and $\hat{\theta}_i$, are also prepared in a similar manner.

Now we consider the dynamics that should be followed by the global and local variables in compliance with the superposition representation in~\eqref{eq:super_ex1}.
Because such a dynamics is not necessarily unique, we can possibly choose a representation that has a desirable property for our controller design.
Indeed, we impose a \emph{hierarchical} structure into the dynamics.
We can show that there exists a hierarchical system
\[
 \left\{
 \begin{array}{lcl}
 \Xi_i: \hs \left[
 \begin{array}{c}
 \dot{\hat{\theta}}_i\\
 \dot{\hat{\omega}}_i
 \end{array}
 \right]
 \hs = \hat{A}_i
 \left[
 \begin{array}{c}
 \hat{\theta}_i\\
 \hat{\omega}_i
 \end{array}
 \right] 
 +B_i\hat{u}_i,\quad i=1,2,3\\
 \Xi_0: \hs \left[
 \begin{array}{c}
 \dot{\hat{\theta}}_0\\
 \dot{\hat{\omega}}_0
 \end{array}
 \right] \hs = \hat{A}_0
 \left[
 \begin{array}{c}
 \hat{\theta}_0\\
 \hat{\omega}_0
 \end{array}
 \right]
 \displaystyle{+\sum_{i=1}^N \hat{R}_i}
 \left[
 \begin{array}{c}
 \hat{\theta}_i\\
 \hat{\omega}_i
 \end{array}
 \right]
 + B_0\hat{u}_0
 \end{array}
 \right.
\]
with certain system matrices such that the original states $\theta_i$ and $\omega_i$ can be reproduced as a superposition of the states for any control inputs as long as the initial condition is consistent.
Although the specific description of the matrices is omitted here, its general derivation is discussed in Sec.~\ref{subsec:hier}.
Block diagrams of the original network system and the hierarchical system are illustrated at the top of Fig.~\ref{fig:hier_sec_ex}.

The hierarchical representation can be interpreted as dynamics in response to multiple local disturbances.
The dynamics $\Xi_0$ stands for a reduced-order global model of the original network system, while the dynamics of $\Xi_1,\Xi_2,\Xi_3$ represent local models.
In this sense, the hierarchical representation decomposes the entire model into global and local models.
At the bottom of Fig.~\ref{fig:hier_sec_ex}, the state trajectories without control in response to a disturbance occurring in the first two clusters are depicted.
Accordingly, the local state $\hat{\omega}_3$ in the third cluster is not excited at all, which is induced from the hierarchical structure.
Moreover, the original state trajectories can be represented as a superposition of the states in the hierarchical system as shown at the bottom of this figure.
In a similar manner, this hierarchical decomposition can reproduce the original state as a superposition of global and local states for any disturbance and any control inputs.

\begin{figure}[t]
\centering
\subfloat[][Hierarchical model decomposition with control inputs.]{\includegraphics[width=.48\linewidth]{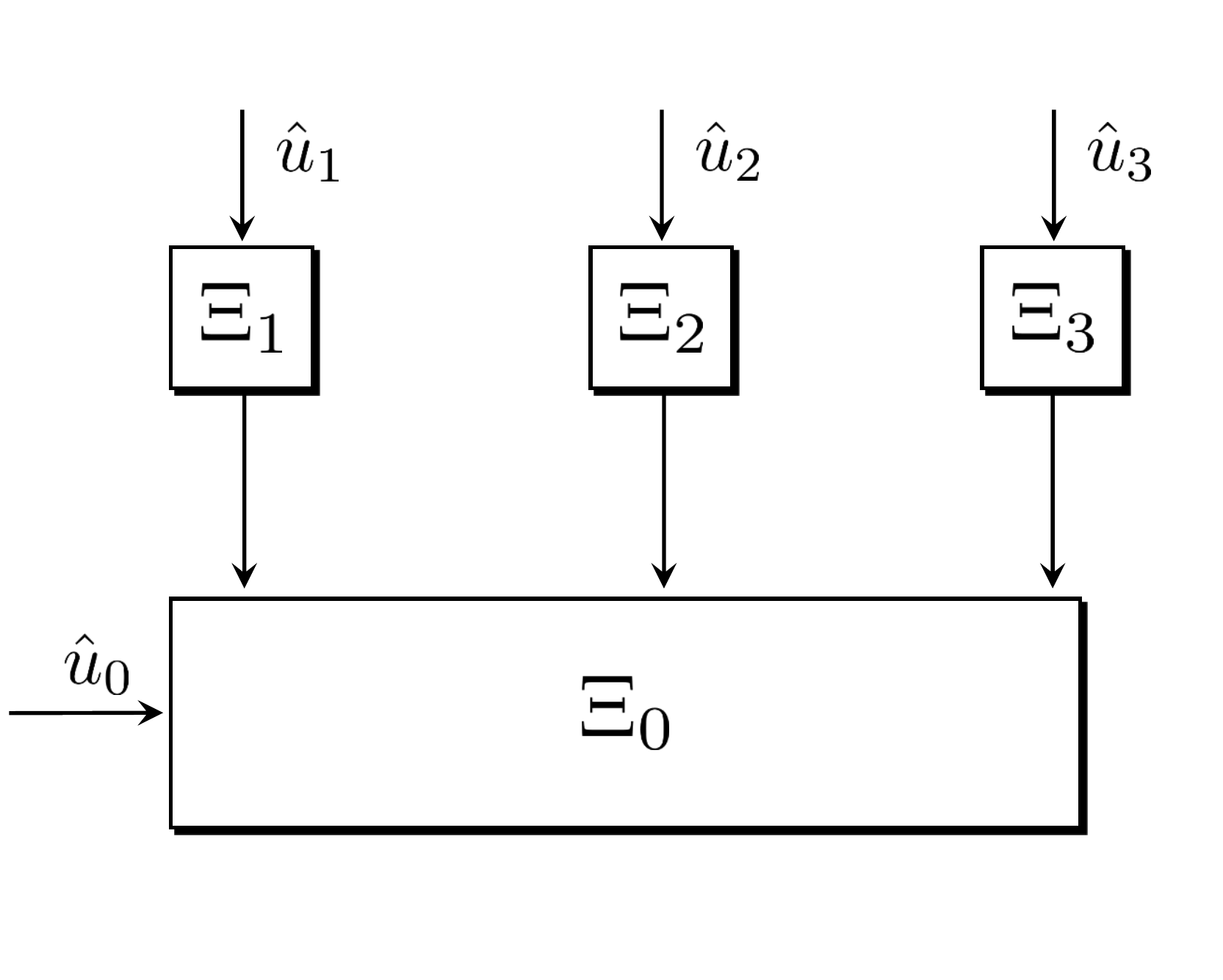}\label{subfig:open}} \quad
\subfloat[][Hierarchical model decomposition under feedback control.]{\includegraphics[width=.48\linewidth]{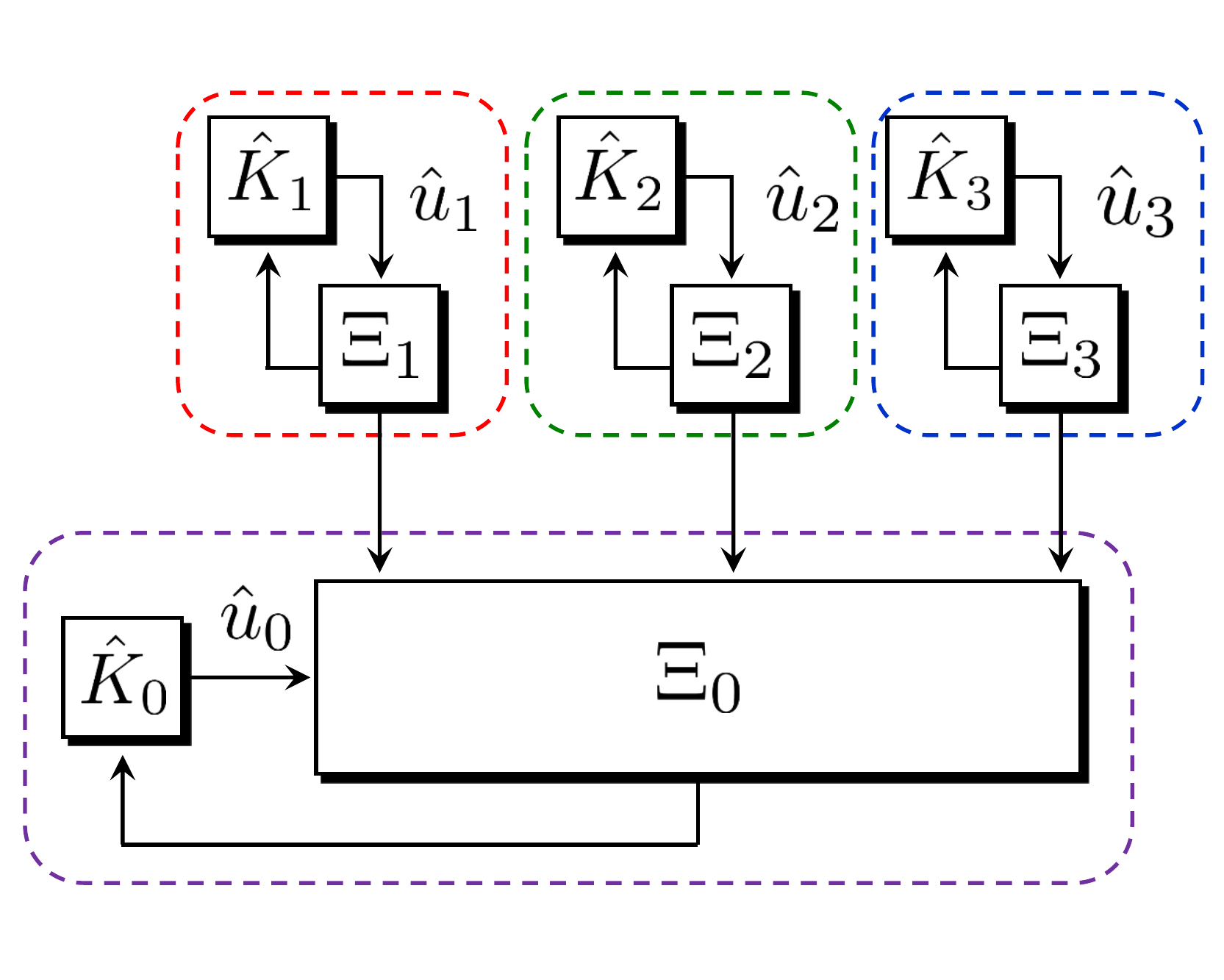}\label{subfig:close}} \quad
\caption[]{Block diagrams of the hierarchical model decomposition with control signals.}
\label{fig:hier_ex}
\end{figure}

Consider utilizing this decomposition for our problem.
A block diagram of $\Xi$ with control inputs is illustrated in Fig.~\ref{subfig:open}.
From the hierarchical structure, there are no feedback paths around the reduced-order models.
Thus we can guarantee stability of the whole system by attaching subcontrollers as in Fig.~\ref{subfig:close} provided that each subcontroller stabilizes the corresponding reduced-order system.
Following this procedure, we can perform distributed design of a glocal controller, where $\mathcal{K}_i$ is given as a set containing stabilizing controllers for $\Xi_i$.

In the rest of this section, we adopt the scheme explained above to the general systems and then present some important properties for control design and implementation.

\subsection{Definition of Hierarchical Model Decomposition}
\label{subsec:hier}
The key idea of the proposed method is to represent the entire network system as a hierarchical system consisting of reduced-order models.
A definition of the hierarchical model, referred to as \emph{hierarchical model decomposition,} is given as follows.
\begin{defin}[Hierarchical Model Decomposition]\label{def:hier}
Consider the hierarchical system $\Xi$ composed of
\begin{equation}\label{eq:hier_sys}
 \left\{
 \begin{array}{lcl}
 \Xi_i: \hs \dot{\xi}_i \hs = \hat{A}_i\xi_i+B_i\hat{u}_i,\quad i=1,\ldots,N\\
 \Xi_0: \hs \dot{\xi}_0 \hs = \hat{A}_0\xi_0+\sum_{i=1}^N \hat{R}_i\xi_i + B_0\hat{u}_0.
 \end{array}
 \right.
\end{equation}
The system $\Xi$ is said to be a hierarchical model decomposition of the clustered interconnected system in~\eqref{eq:cl_sys} if
\begin{equation}\label{eq:super}
 \textstyle{x(t) = \sum_{i=1}^N P_i \xi_i(t) + P_0 \xi_0(t),\quad \forall t \geq 0}
\end{equation}
holds for arbitrary initial conditions and control inputs provided that $x(0) = \sum_{i=1}^N P_i \xi_i(0) + P_0 \xi_0(0)$.
\end{defin}

If we can obtain a hierarchical model decomposition~\eqref{eq:hier_sys},
then distributed design can be achieved.
The technical questions related to the decomposition are as follows:
\begin{enumerate}
\item Does there exist a hierarchical model decomposition for the given system and clusters?
\item How to obtain a specific representation of a hierarchical model decomposition if it exists?
\item How to implement the designed controller preserving the cascade structure?
\end{enumerate}
Those questions are addressed in the remainder of this section.

\subsection{Existence Condition and Implicit Representation}
\label{subsec:ex_hier}

We first give a necessary and sufficient condition on existence of hierarchical model decomposition from the viewpoint of controllable subspaces in terms of given clusters.
\begin{theorem}[Existence Condition]\label{thm:ex}
Under Assumption~\ref{assum:identical},
a hierarchical model decomposition of~\eqref{eq:cl_sys} exists if and only if the condition
\begin{subnumcases}
{}
 \mathcal{R}(A,P_i) \subset \im{P_i}+\im{P_0},\quad i=1,\ldots,N \label{eq:excond1}\\
 \mathcal{R}(A,P_0) \subset \im{P_0} \label{eq:excond2}
 \end{subnumcases}
holds.
\end{theorem}

The condition in Theorem~\ref{thm:ex} can be interpreted as follows.
Consider possible effects of a disturbance and a control input injected into the $i$th cluster.
Condition~\eqref{eq:excond1} indicates that state trajectories excited by the inputs are restricted in the space of the right hand side of~\eqref{eq:excond1}.
Consequently, the behavior in response to the disturbance and control input can be represented as a sum of global and local behaviors.
Condition~\eqref{eq:excond2} implies that $\im{P_0}$ is an invariant subspace of $A$.
Then the global control input $\hat{u}_0$ in~\eqref{eq:cl_sys} can excite only global interarea behaviors restricted to $\im{P_0}$.
The relationship between the controllable subspaces and the image spaces of the matrices $\{P_i\}_{i=0}^N$ induces~\eqref{eq:hier_sys}, which has the special structure illustrated in Fig.~\ref{subfig:open}.

Based on Theorem~\ref{thm:ex}, we derive a necessary and sufficient condition that $\Xi$ must satisfy to be a hierarchical model decomposition.
\begin{theorem}[Implicit Representation]\label{thm:rep}
Under Assumption~\ref{assum:identical}, the system $\Xi$ in~\eqref{eq:hier_sys} is a hierarchical model decomposition of~\eqref{eq:cl_sys} if and only if the condition
\begin{equation}\label{eq:repcon}
 \left\{
 \begin{array}{l}
 AP_i-P_0\hat{R}_i-P_i\hat{A}_i=0,\quad i=1,\ldots,N\\
 AP_0-P_0\hat{A}_0 = 0
 \end{array}
 \right.
\end{equation}
holds.
\end{theorem}

Theorem~\ref{thm:rep} gives an implicit representation of all hierarchical model decompositions through the linear matrix equations~\eqref{eq:repcon}, which can readily be solved.

\emph{Remark:}
\emph{Retrofit control}~\cite{Ishizaki2018Retrofit,Ishizaki2019Modularity} proposes a distributed design method of decentralized controllers through the following hierarchical system representation
\begin{equation}\label{eq:sys_retro}
 \left\{
 \begin{array}{cl}
 \dot{\xi}_i \hs = A_i\xi_i+B_i\hat{u}_i,\quad i=1,\ldots,N\\
 \dot{\xi}_0 \hs = A\xi_0+\sum_{i=1}^N (AP_i-P_iA_i)\xi_i.
 \end{array}
 \right.
\end{equation}
The representation~\eqref{eq:sys_retro} can be obtained by choosing
\[
 N=N_0,\quad \mathcal{I}_i=\{i\},\quad \hat{A}_i=A_i,\quad \hat{u}_0=0
\]
in the framework of this paper.
Under this choice, the existence condition in Theorem~\ref{thm:ex} always holds since $\im{P_0}$ contains the entire space.
Thus, any network systems can be represented in the form~\eqref{eq:sys_retro}, through which distributed design of local subcontrollers can be carried out.
In this sense, hierarchical model decomposition proposed in this paper can be interpreted as an extension of hierarchical state-space expansion in~\cite{Ishizaki2018Retrofit}.

\subsection{Controller Implementation}
\label{subsec:imp}

We consider implementation of subcontrollers designed based on a hierarchical model decomposition.
Let $\hat{K}_0,\ldots,\hat{K}_N$ be stabilizing controllers for the subsystems in $\Xi$ obtained through Theorem~\ref{thm:rep}.
Then the closed-loop systems
\begin{equation}\label{eq:ui}
\left\{
 \begin{array}{cl}
 \dot{\xi}_i \hs = \hat{A}_i\xi_i+B_i\hat{u}_i\\
 \hat{u}_i \hs = \hat{K}_i(C_i\xi_i)
 \end{array},\quad i=0,\ldots,N
\right.
\end{equation}
are internally stable.
Consequently, the entire network system with those control inputs is also stabilized from the cascade structure of $\Xi$ and the identity~\eqref{eq:super}.
However, since the virtual variable $\xi_i$ is inaccessible, the control inputs in~\eqref{eq:ui} cannot be created straightforwardly.
The aim of this subsection is to develop an implementation method to circumvent this problem.

First, notice that the global measurement signal $y_0$ can be used instead of $C_0\xi_0$ in~\eqref{eq:ui} for stabilizing $\Xi_0$, the downstream part of $\Xi$.
Suppose that $\hat{K}_0$ is a linear controller and $y_0$ is used as a feedback signal to $\hat{K}_0$.
Then the control input can be represented by
\[
 \textstyle{\hat{u}_0=\hat{K}_0(y_0)=\hat{K}_0(C_0\xi_0)+\hat{K}_0\left(C_0P_0^{\sf T}\sum_{i=1}^N P_i \xi_i\right).}
\]
From the perspective of the closed-loop system composed of $\Xi_0$ and $\hat{K}_0$, the second term can be regarded as an external input signal from the upstream parts.
Because there is no feedback path from $\Xi_0$ to $\Xi_i$ for $i=1,\ldots,N$, the stability of the downstream part can be guaranteed even if the original global measurement signal $y_0$ is used as an alternative to $C_0\xi_0$.
For this reason, we henceforth confine attention only to the upstream parts associated with $\xi_1,\ldots,\xi_N$.

\begin{figure}[t]
\centering
\includegraphics[width=.98\linewidth]{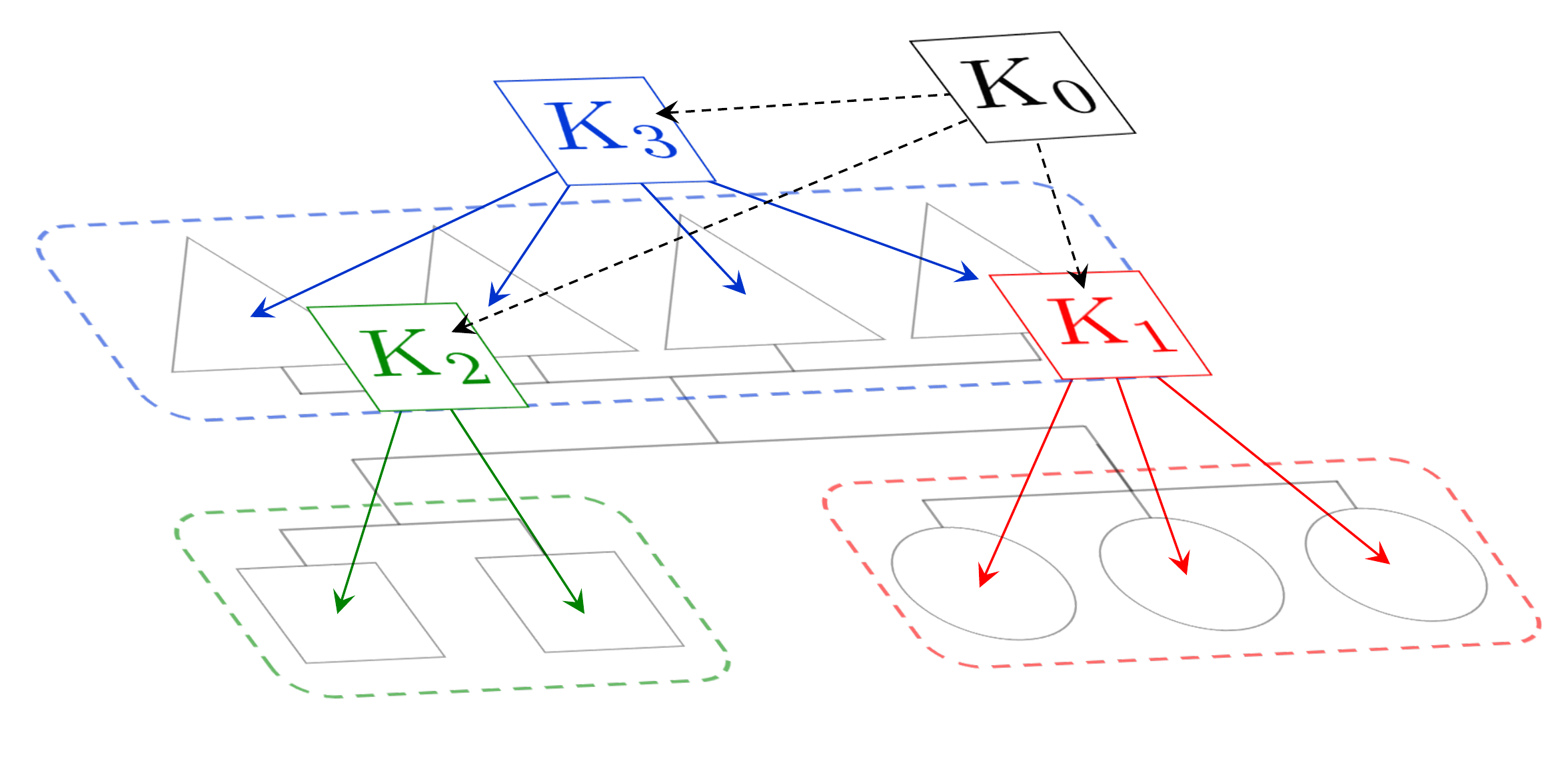}
\caption{The information structure of the glocal controller to be designed, where local subcontrollers $K_1,K_2,K_3$ receive the global control input from the global subcontroller $K_0$ and all local subcontrollers do not directly communicate with each other.}
\label{fig:structure}
\end{figure}

The idea for implementation is to estimate $C_i\xi_i$ for $i=1,\ldots,N$ using functional observers~\cite{Darouach2000Existence,Fortmann1972Design}.
A functional observer is a state observer producing a signal that tracks partial states of interest.
For designing a functional observer, we consider the particular information structure given by
\[
 \hat{u}_0=K_0(y_0),\quad \hat{u}_i=K_i(y_i,\hat{u}_0),\quad i=1,\ldots,N.
\]
This structure is illustrated in Fig.~\ref{fig:structure}, where the whole controller has a star topology.
The global subcontroller transmits its control input to all local subcontrollers, while the local subcontrollers do not directly communicate with each other.

The following theorem claims that the entire network system can be stabilized by combining functional observers of $C_1\xi_1,\ldots,C_N\xi_N$ with the subcontrollers~\eqref{eq:ui}.
\begin{theorem}[Stabilization through Functional Observers]\label{thm:linob}
Assume that $\Xi$ is a hierarchical model decomposition of~\eqref{eq:cl_sys} and that
the dynamical systems
\begin{equation}\label{eq:linob}
 \Phi_i: \left\{
 \begin{array}{cl}
 \dot{\phi}_i \hs = \mathbf{A}_i\phi_i+\mathbf{B}_{i}\hat{u}_i+\mathbf{D}_{i}\hat{u}_0+\mathbf{E}_{i}y_i\\
 \psi_i \hs = \mathbf{C}_i\phi_i+\mathbf{F}_iy_i
 \end{array}
 \right.
\end{equation}
for $i=1,\ldots,N$ are functional observers of $C_i\xi_i$, i.e.,
\[
 \textstyle{\lim_{t \rightarrow +\infty}\left(C_i\xi_i(t)-\psi_i(t)\right)=0}
\]
holds for any initial condition and inputs.
Then the glocal controller composed of
\begin{equation}\label{eq:ctrls}
 \hat{u}_0=\hat{K}_0y_0,\quad \hat{u}_i=\hat{K}_i\psi_i,\quad i=1,\ldots,N
\end{equation}
with~\eqref{eq:linob} stabilizes the clustered interconnected system~\eqref{eq:cl_sys}
for any $\hat{K}_0,\ldots,\hat{K}_N$ such that the closed-loop systems~\eqref{eq:ui} are internally stable.
\end{theorem}

Theorem~\ref{thm:linob} implies that the stability of the original system can be guaranteed using the estimated signal $\psi_i$ instead of $C_i\xi_i$ itself.

We next provide a specific functional observer for~\eqref{eq:linob}.
For simplicity, we consider the case where the interaction signal $v_i$ can be measured by the $i$th local subcontroller in addition to the local measurement signal $y_i$ for $i=1,\ldots,N$.
\begin{prop}\label{thm:exlinob}
Assume that $A_i$ and $\hat{A}_i$ are stable.
Then
\begin{equation}\label{eq:exlinob}
 \Phi_i: \left\{
 \begin{array}{cl}
  \dot{\phi}_i & \hspace{-3mm} = \hat{A}_i \phi_i + (A_i-\hat{A}_i)\hat{x}_i+L_iv_i+P_i^{\sf T}P_0B_0\hat{u}_0\\
  \dot{\hat{x}}_i & \hspace{-3mm} = A_i\hat{x}_i +B_i\hat{u}_i+L_iv_i+P_i^{\sf T}P_0B_0\hat{u}_0\\
  \psi_i &\hspace{-3mm} = -C_i\phi_i + y_i
 \end{array}
 \right.
\end{equation}
is a functional observer of $C_i\xi_i$.
\end{prop}

The idea to construct the observer in Proposition~\ref{thm:exlinob} is as follows.
Notice that $y_i=C_i\xi_i+C_iP_i^{\sf T}P_0\xi_0$ implies that if we use $y_i$ as a feedback signal then the cascade structure of $\Xi$ is collapsed because of the arising feedback path from $\Xi_0$ to $\Xi_i$.
To extract $C_i\xi_i$ alone, consider estimating $P_i^{\sf T}P_0\xi_0$ using $v_i,\hat{u}_0,$ and an estimated local state $\hat{x}_i$ generated through the second differential equation in~\eqref{eq:exlinob}.
Because the dynamics of $P_i^{\sf T}P_0\xi_0$ can be represented by the first differential equation in~\eqref{eq:exlinob},
we use $\phi_i$ as a replacement of $P_i^{\sf T}P_0\xi_0$, which induces $\psi_i=y_i-C_i\phi_i$ as an estimation of $C_i\xi_i=y_i-C_iP_i^{\sf T}P_0\xi_0$.
We stress that the requirement for each functional observer is independent of the other subsystems, subcontrollers, and functional observers.
Hence, design of functional observers does not hinder distributed design of the glocal controller.

Our solution to Problem~\ref{prob:1} can be summarized by the following procedure.
\begin{enumerate}
\item Identify existence of a hierarchical model decomposition for the given clustered interconnected system based on Theorem~\ref{thm:ex}.
\item Construct a hierarchical model decomposition based on Theorem~\ref{thm:rep}.
\item Design $\mathcal{K}_i$ to be the set whose elements are the controllers composed of the functional observer~\eqref{eq:exlinob} and an internal controller that stabilizes~\eqref{eq:ui} for $i=0,\ldots,N$.
\end{enumerate}

In the above procedure, it is assumed that the clusters are given in advance and satisfy the existence condition in Theorem~\ref{thm:ex}.
In Sec.~\ref{sec:clust}, we develop a clustering method that provides a solution to Problem~\ref{prob:2} based on the results in this section.
Moreover, an extension of the proposed approach to the case where the existence condition is \emph{not} satisfied is discussed in Sec.~\ref{sec:rob}.

\section{Clustering Method}
\label{sec:clust}

\subsection{Clustering Algorithm}

This section addresses Problem~\ref{prob:2}.
For the given interconnected system composed of~\eqref{eq:component_k} and the interaction~\eqref{eq:inter_components}, we find a cluster set for which there exists a hierarchical model decomposition of the clustered interconnected system~\eqref{eq:cl_sys}.
Before proceeding, we state an alternative assumption to Assumption~\ref{assum:identical}.
Instead, the following assumption is made.
\begin{assum}\label{assum:identical_clustering}
The input and output matrices of all components are identical, that is, $B_{[k]}=B_{[l]}$ and $C_{[k]}=C_{[l]}$ for any $k,l =1,\ldots,N_0$.
\end{assum}

It is clear that if Assumption~\ref{assum:identical_clustering} holds then Assumption~\ref{assum:identical} is satisfied for any clusters.
When Assumption~\ref{assum:identical_clustering} does not hold, it suffices to modify the global signals as explained in Sec.~\ref{subsec:sys} after the clustering.

The desirable clusters are characterized through~\eqref{eq:excond1} and~\eqref{eq:excond2}.
Since cluster sets that fulfill the conditions are not unique, we have to make a criterion to choose.
A trivial cluster set is given by $\mathcal{I}_i=\{i\}$ for $i=1,\ldots,N_0$.
As mentioned in the remark in Sec.~\ref{subsec:ex_hier}, this choice does not reduce the complexity of designing a global subcontroller at all.
Noticing that the number of the clusters is maximized by choosing the trivial cluster set,
we aim at minimizing the number of clusters.
An important observation is that the condition~\eqref{eq:excond1} becomes a sufficient condition of~\eqref{eq:excond2} under a mild condition.
\begin{prop}\label{lem:suf}
Assume that $(A,\{P_i\}_{i=0}^N)$ satisfies
\begin{equation}\label{eq:sing}
 \im{P_0^{(i)}}\subset \mathcal{R}(A,[P_1\ \cdots P_{i-1}\ P_{i+1}\ \cdots\ P_N]),\quad i=1,\ldots,N
\end{equation}
where $P_0^{(i)} := \pi_{\im{P_i}} P_0$
with $\pi_{\im{P_i}}:=P_iP_i^{\sf T}$, a projection matrix onto $\im{P_i}$.
Then, if~\eqref{eq:excond1} holds,~\eqref{eq:excond2} holds as well.
\end{prop}

Note that $P_0^{(i)}$ satisfies the relationship
\[
 \im{P_0^{(i)}} = \im{P_i} \cap \im{P_0},
\]
which suggests that $\im{P_0^{(i)}}$ contains synchronized states in the $i$th cluster.
In Proposition~\ref{lem:suf}, the condition~\eqref{eq:sing} means that the subspace $\im{P_0^{(i)}}$ is reachable from one of the other clusters.
Since this condition is not very restrictive,
we first develop an algorithm producing clusters that satisfy~\eqref{eq:excond1} disregarding~\eqref{eq:excond2},
and subsequently we extend the algorithm to satisfy~\eqref{eq:sing} as well.

\begin{figure}[t]
\centering
\subfloat[][The clusters at the $\tau$th step.]{\includegraphics[width=.48\linewidth]{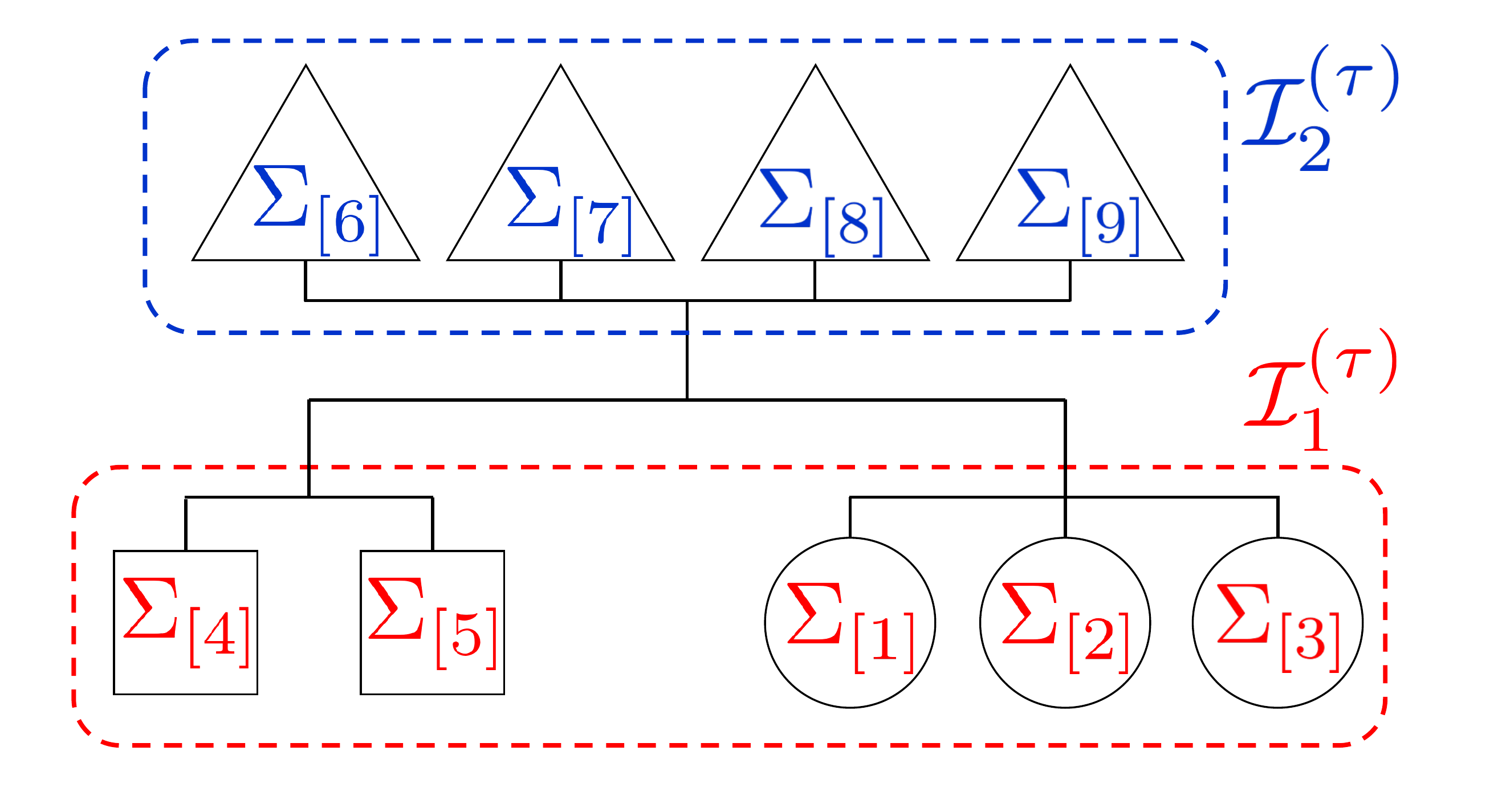}\label{subfig:cluster1}} \quad
\subfloat[][The clusters at the $(\tau+1)$th step.]{\includegraphics[width=.48\linewidth]{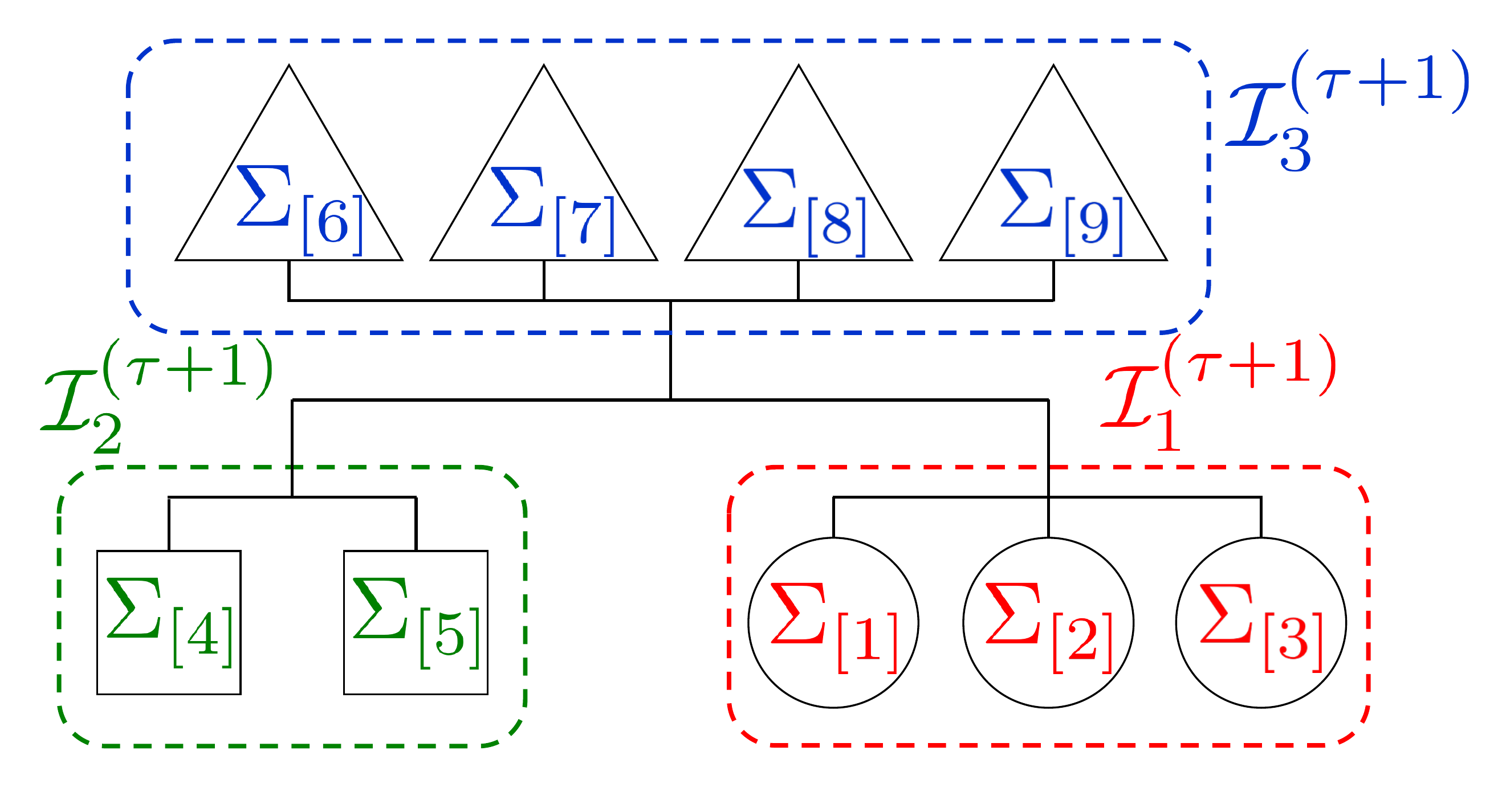}\label{subfig:cluster2}} \quad
\caption[]{An illustrative example of the clusters during the algorithm.}
\label{fig:clustering}
\end{figure}

We exemplify our proposed algorithm through the motivating example in Sec.~\ref{subsec:ex}, where the desired clusters are given by~\eqref{eq:clu_ex}.
The proposed algorithm is based on a greedy approach.
We begin with an initial cluster set, partition one of the clusters such that~\eqref{eq:excond1} is satisfied, and repeat this process until all partitioned clusters satisfy~\eqref{eq:excond1}.
Suppose that the current step is the $\tau$th step at which the temporary cluster set is given as the one in Fig.~\ref{subfig:cluster1}.
We check if the clusters satisfy~\eqref{eq:excond1}.
Observe that
\[
 P_2^{(\tau)}=\left[
 \begin{array}{c}
 0_{10\times 8}\\
  I_{8}
 \end{array}  
 \right],\quad
 P_0^{(\tau)}=\left[
 \begin{array}{cc}
 \mathds{1}_5\otimes I_2 & 0\\
 0 & \mathds{1}_4\otimes I_2
 \end{array}
 \right]
\]
and the controllable subspace from the second cluster
\begin{equation}\label{eq:cl_div}
 \mathcal{R}(A,P_2^{(\tau)})= \im{P^{(\tau)}_2}\oplus \im{\left[
 \begin{array}{c}
 \mathds{1}_3\otimes I_2\\
 0_{4\times2}\\
 0_{8\times2}
 \end{array}
 \right]}
 \oplus \im{\left[
 \begin{array}{c}
 0_{6\times2}\\
 \mathds{1}_2\otimes I_2\\
 0_{8\times2}
 \end{array}
 \right]}.
\end{equation}
Thus~\eqref{eq:excond1} does not hold.
Accordingly, we choose $\mathcal{I}^{(\tau)}_1$ to be partitioned into multiple clusters at the next step.
From~\eqref{eq:cl_div}, the components $\Sigma_{[k]}$ for $k=1,2,3$ can be lumped together, and $\Sigma_{[4]}$ and $\Sigma_{[5]}$ can be lumped together as well.
This procedure can systematically be performed by comparing rows of the controllability matrix.
This partition results in the clusters illustrated by Fig.~\ref{subfig:cluster2} at the $(\tau+1)$th step.
Then we can terminate the algorithm by confirming that the condition~\eqref{eq:excond1} is satisfied for $i=1,2,3$.
Note that~\eqref{eq:sing} is satisfied in this case and hence~\eqref{eq:excond2} also holds.

The detailed description of the clustering algorithm is shown in Algorithm~1,
where $f_{P_i}(\mathcal{I}_i)$ generates the matrix $P_i$ according to~\eqref{eq:Ps},
$f_{\mathbf{R}}(A,P_i)$ generates the controllability matrix $\mathbf{R}_i$ with respect to the pair $(A,P_i)$,
${\rm par}(\mathbf{R}_i,P_i)$ generates the minimal clusters that satisfy the condition~\eqref{eq:excond1} for $P_i$ based on $\mathbf{R}_i$ through the optimization problem
\begin{equation}\label{eq:opt}
 \displaystyle{\min_{\{\mathcal{I}_j\}_{j=1}^{N}}} N\ 
 {\rm s.t.}\ \pi_{\im{P_i}^{\perp}}\,\mathcal{R}(A,P_i) \subset \pi_{\im{P_i}^{\perp}}\,\im{f_{P_0}(\{\mathcal{I}_j\}_{j=1}^N)},
\end{equation}
with $\pi_{\im{P_i}^{\perp}}:=I-P_iP_i^{\sf T}$, a projection matrix onto the orthogonal subspace of $\im{P_i}$, with which the state of the components in the $i$th cluster is disregarded,
and $f_{P_0}(\{\mathcal{I}_i\})$ generates the matrix $P_0$ according to~\eqref{eq:Ps}.
The subproblem, which can be regarded as a greedy part, can be solved by comparing rows of $\pi_{\im{P_i}^{\perp}}\,\mathbf{R}_i$ as mentioned above.

Clearly, Algorithm~1 produces a cluster set that satisfies~\eqref{eq:excond1}.
It should be remarked that if we use the single largest cluster $\mathcal{I}^{(0)}_1=\{1,\ldots,N_0\}$
as the initial cluster set then the condition~\eqref{eq:excond1} holds for any systems.
Thus Algorithm~1 is immediately terminated with the initial cluster set, not producing any beneficial cluster set.
For this reason, the initial cluster set must be set to have at least two clusters.

\begin{algorithm}[th]
\caption{Clustering Algorithm}
\begin{algorithmic}[1]
\REQUIRE{$A,\{\mathcal{I}_i^{(0)}\}_{i=1}^{N^{(0)}}$}
\ENSURE{$\{\mathcal{I}_i\}_{i=1}^N$}
\STATE $\tau \leftarrow 0$
\REPEAT
\STATE $\tau \leftarrow \tau+1$
\FOR{$i=1,\ldots,N^{(\tau-1)}$}
\STATE $P_i^{(\tau-1)}\leftarrow f_{P_i}(\mathcal{I}^{(\tau-1)}_i)$
\STATE $\mathbf{R}^{(\tau-1)}_i \leftarrow f_{\mathbf{R}}(A,P_i^{(\tau-1)})$
\IF {the condition~\eqref{eq:excond1} is not satisfied for $i$}
\STATE $\{\mathcal{I}\}_{i=1}^{N^{(\tau)}} \leftarrow {\rm par}(\mathbf{R}_i^{(\tau-1)},P_i^{(\tau-1)})$
\STATE \textbf{break}
\ENDIF
\ENDFOR
\UNTIL{$\{\mathcal{I}^{(\tau)}_i\}_{i=1}^{N^{(\tau)}} = \{\mathcal{I}_i^{(\tau-1)}\}_{i=1}^{N^{(\tau-1)}}$}
\STATE $\{\mathcal{I}_i\}_{i=1}^{N} = \{\mathcal{I}_i^{(\tau)}\}_{i=1}^{N^{(\tau)}}$
\end{algorithmic}
\end{algorithm}

\subsection{Algorithm Property}

We show the optimality of Algorithm~1.
The following notion is needed.
\begin{defin}[Partition of Clusters]\label{def:per}
A cluster set $\{\mathcal{I}_i\}_{i=1}^N$ is said to be a \emph{partition} of $\{\mathcal{I}_{i'}\}_{i'=1}^{N'}$ when
for any $i\in\{1,\ldots,N\}$ there exists $i' \in \{1,\ldots,N'\}$ such that $I_i\subset I_{i'}$.
\end{defin}

Let $\mathfrak{F}(\{\mathcal{I}_i\}_{i=1}^N)$ be the family of all cluster sets that are partitions of $\{\mathcal{I}_i\}_{i=1}^N$.
Intuitively, $\mathfrak{F}(\{\mathcal{I}_i\}_{i=1}^N)$ contains all cluster sets that can be produced from the original cluster set $\{\mathcal{I}_i\}_{i=1}^N$ by partitioning some of the clusters.
Moreover, let $\mathfrak{G}(\{\mathcal{I}_i\}_{i=1}^N) \subset \mathfrak{F}(\{\mathcal{I}_i\}_{i=1}^N)$ denote the family of all cluster sets in $\mathfrak{F}(\{\mathcal{I}_i\}_{i=1}^N)$ such that~\eqref{eq:excond1} is satisfied.
The following theorem shows minimality of the resulting cluster set.
\begin{theorem}[Algorithm Property]\label{thm:subopt}
The cluster set produced by Algorithm~1 is the minimum cluster set in $\mathfrak{G}(\{\mathcal{I}^{(0)}_i\}_{i=1}^{N^{(0)}})$.
\end{theorem}

Essentially, Theorem~\ref{thm:subopt} is a direct consequence of the following lemma.
\begin{lem}\label{lem:clusters}
The relation
\[
 \mathfrak{G}(\{\mathcal{I}^{(0)}_i\}_{i=1}^{N^{(0)}}) = \mathfrak{G}(\{\mathcal{I}^{(\tau)}_i\}_{i=1}^{N^{(\tau)}})
\]
holds for any $\tau \geq 0$.
\end{lem}

Lemma~\ref{lem:clusters} implies that Algorithm~1 preserves the admissible clusters.
Therefore, we can guarantee minimality of the resulting cluster set in $\mathfrak{G}(\{\mathcal{I}^{(0)}_i\}_{i=1}^{N^{(0)}})$.

\subsection{Extended Clustering Algorithm}

We next extend Algorithm~1 to the case when the resulting cluster set does not satisfy~\eqref{eq:excond2}.
In this case,~\eqref{eq:sing} is neither satisfied. 
The idea for the extension is to apply Algorithm~1 only to a subset of clusters that do not satisfy~\eqref{eq:sing} inspired by the following proposition.
\begin{prop}\label{prop:invariance}
Let $\{\mathcal{I}_i\}_{i=1}^N$ be a cluster set such that~\eqref{eq:excond1} is satisfied for any $i\in\{1,\ldots,N\}$ and~\eqref{eq:sing} is not satisfied for $j \in \mathcal{J} \subset \{1,\ldots,N\}$.
Then for any cluster set $\{\mathcal{I}'_{j'}\}_{j'\in\mathcal{J}'}$ in $\mathfrak{F}(\{\mathcal{I}_j\}_{j\in\mathcal{J}})$,
the following cluster set
\[
 \{\mathcal{I}_i\}_{i\notin \mathcal{J}}\cup \{\mathcal{I}_{j'}\}_{j'\in \mathcal{J}'}
\]
which is obtained by partitioning $\{\mathcal{I}_j\}_{j \in \mathcal{J}}$ into $\{\mathcal{I}'_{j'}\}_{j'\in \mathcal{J}'}$,
satisfies~\eqref{eq:excond1} as well as~\eqref{eq:sing} for $i\notin \mathcal{J}$.
\end{prop}

Proposition~\ref{prop:invariance} implies that, once~\eqref{eq:excond1} and~\eqref{eq:sing} are satisfied for some clusters, this property is preserved even under partition of the other clusters.
Thanks to Proposition~\ref{prop:invariance}, we can reduce the clustering problem into a subproblem for the subclusters that do not satisfy~\eqref{eq:sing}.

The proposed clustering algorithm is described in Algorithm~2.
As mentioned above, the idea is to apply Algorithm~1 repeatedly with an initial partition of the clusters such that~\eqref{eq:sing} is not satisfied.
The resulting clusters obviously satisfy the conditions in Theorem~\ref{thm:ex} from Propositions~\ref{lem:suf} and~\ref{prop:invariance}.
\begin{prop}\label{prop:ex_clustering}
Consider the clusters produced by Algorithm~2.
Under Assumption~\ref{assum:identical_clustering}, there exists a hierarchical model decomposition of the resulting clustered interconnected system.
\end{prop}


\begin{algorithm}[th]
\caption{Extended Clustering Algorithm}
\begin{algorithmic}[1]
\REQUIRE{$A,\{\mathcal{I}_i\}^{(0)}$}
\ENSURE{$\{\mathcal{I}_i\}_{i=1}^N$}
\STATE $\tau' \leftarrow 0$
\STATE $\{\mathcal{I}_i\}^{(\tau')} \leftarrow \{\mathcal{I}_i\}^{(0)}$
\REPEAT
\STATE $\tau' \leftarrow \tau'+1$
\STATE $\{\mathcal{I}_i\}^{(\tau')} \leftarrow$ Algorithm~1 with $\{\mathcal{I}_i\}^{(\tau'-1)}$
\IF{\eqref{eq:sing} is not satisfied for $j\in\mathcal{J}\subset \{1,\ldots,N\}$}
\STATE provide $\{\mathcal{I'}_{j'}\}_{j'\in \mathcal{J}'}$ in $\mathfrak{F}(\{\mathcal{I}_j\}_{j\in \mathcal{J}})$
\STATE $\{\mathcal{I}_i\}^{(\tau')} \leftarrow \{\mathcal{I}_i\}_{i\notin \mathcal{J}} \cup \{\mathcal{I'}_{j'}\}_{j'\in \mathcal{J}'}$
\STATE \textbf{break}
\ENDIF
\UNTIL{the condition~\eqref{eq:sing} is satisfied for all clusters}
\STATE $\{\mathcal{I}_i\}_{i=1}^{N} = \{\mathcal{I}_i\}^{(\tau')}$
\end{algorithmic}
\end{algorithm}

\section{Extension to Indecomposable Systems}
\label{sec:rob}

In the framework proposed in Sec.~\ref{sec:hier}, the state $x$ is needed to be perfectly represented as the superposition of $\xi_0,\xi_1,\ldots,\xi_N$ without error as in~\eqref{eq:super}.
This section extends the proposed method to systems for which no exact hierarchical model decompositions exist.

\subsection{Motivating Example Revisited}

\begin{figure}[t]
\centering
\includegraphics[width=0.98\linewidth]{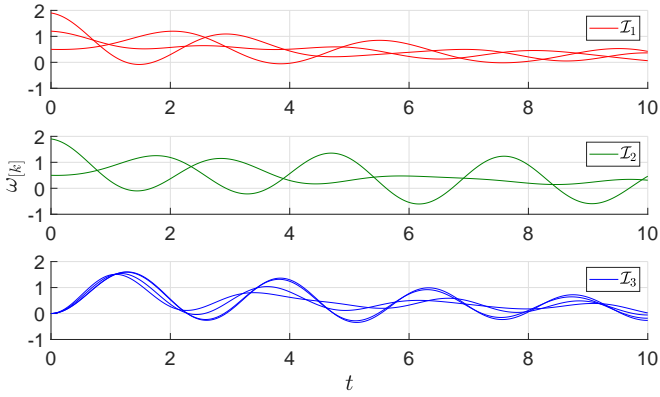}
\caption{Free response of the second-order system with slightly perturbed model parameters.}
\label{fig:free_rob}
\end{figure}

Consider the motivating example again..
We assume that the model parameters are perturbed away from their nominal values.
Specifically, let $m_{[k]}=(1+\delta_{m,[k]})\overline{m}_{[k]}$ and $d_{[k]}=(1+\delta_{d,[k]})\overline{d}_{[k]}$
where $(\overline{m}_{[k]},\overline{d}_{[k]})$ are the nominal values given in~\eqref{eq:para_nominal} and $(\delta_{m,[k]},\delta_{d,[k]})$ are independently and randomly generated scalars following the uniform distribution with the interval $[-0.2,0.2]$.
For this system, there exist no hierarchical model decompositions.
However, as depicted by Fig.~\ref{fig:free_rob}, it seems like the system behavior is similar to that in the preceding case of Fig.~\ref{fig:hier_sec_ex}.
The bottom of Fig.~\ref{fig:free_rob} suggests that the state in $\mathcal{I}_3$ can be approximated by the global behavior with a small error.
This example motivates us to extend the proposed method by admitting approximation error in the hierarchical model decomposition.

\subsection{Robust Hierarchical Model Decomposition}

\begin{figure}[t]
\centering
\includegraphics[width=0.98\linewidth]{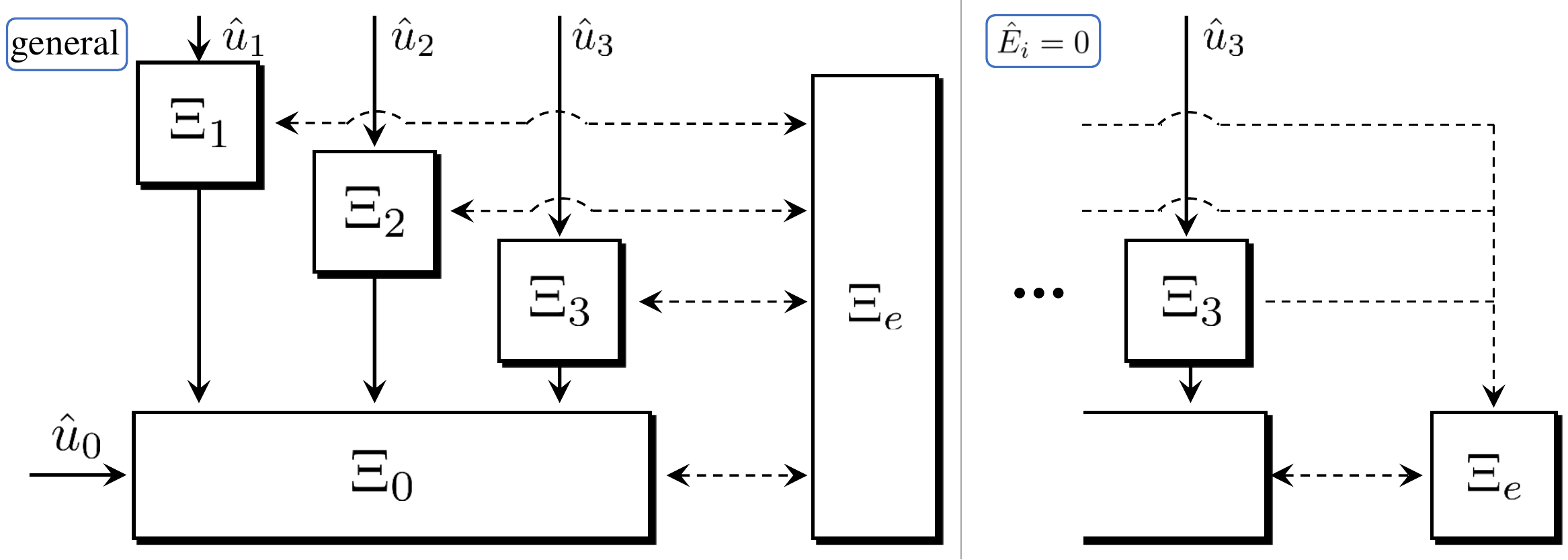}
\caption{Block diagrams of the hierarchical model decomposition with error dynamics in~\eqref{eq:hier_e} when $N=3$.
Left: the general case. Right: the case under the choice~\eqref{eq:Eiz} where the system has a hierarchical structure including the error dynamics.
}
\label{fig:hier_e}
\end{figure}

To handle approximation error, we consider the following system
\begin{equation}\label{eq:hier_e}
 \left\{
 \begin{array}{cl}
 \dot{\xi}_i \hs = \hat{A}_i\xi_i+\hat{E}_ie+B_i\hat{u}_i,\quad i=1,\ldots,N\\
 \dot{\xi}_0 \hs = \hat{A}_0\xi_0+\sum_{i=1}^N \hat{R}_i\xi_i +\hat{E}_0e + B_0\hat{u}_0\\
 \dot{e} \hs =  \hat{A}_ee+\hat{F}_0\xi_0+\sum_{i=1}^N\hat{F}_i\xi_i,
 \end{array}
 \right.
\end{equation}
which is obtained by introducing the error signal $e$ and interaction between $e$ and $\xi_i$ for $i=0,1,\ldots,N$ into $\Xi$ in~\eqref{eq:hier_sys}.
From simple calculation, it turns out that
\[
 \textstyle{x(t)=\sum_{i=1}^N P_i \xi_i(t)+P_0\xi_0(t)+e(t),\quad \forall t\geq 0}
\]
holds for any control inputs and initial states provided that $x(0)=\sum_{i=1}^N P_i \xi_i(0)+P_0\xi_0(0)+e(0)$ if and only if
\begin{equation}\label{eq:paras_e}
\begin{array}{l}
 \hat{A}_e=A-P_0\hat{E}_0-\sum_{i=1}^NP_i\hat{E}_i,\quad \hat{F}_0=AP_0-P_0\hat{A}_0,\\
 \hat{F}_i=AP_i-P_i\hat{A}_i-P_0\hat{R}_i,\quad i=1,\ldots,N
\end{array}
\end{equation}
where $\hat{A}_0,\hat{A}_i,\hat{R}_i,\hat{E}_0,\hat{E}_i$ are free parameters.
The block diagram of the system~\eqref{eq:hier_e} for the case $N=3$ is depicted in the left of Fig.~\ref{fig:hier_e} where $\Xi_e$ represents the dynamics of $e$.
Because there exist feedback paths from $\Xi_e$ to $\Xi_0$ and $\Xi_i$ for $i=1,\ldots,N$ as shown in this figure, the system~\eqref{eq:hier_e} no longer has a cascade structure.
Hence, the entire stability cannot be guaranteed even if we attach subcontrollers each of which stabilizes the corresponding subloop.

An important observation from~\eqref{eq:hier_e} is that the hierarchical cascade structure can be recovered by choosing the free parameters appropriately.
In particular, when the condition
\begin{equation}\label{eq:Eiz}
 \hat{E}_i=0,\quad \forall i=1,\ldots,N
\end{equation}
or
\[
 \hat{E}_0=0\ {\rm and}\ \hat{F}_0=0
\]
is satisfied, the cascade structure is preserved.
In the former case, there are no signals flowing from the error dynamics $\Xi_e$ to the upstream parts $\Xi_1,\ldots,\Xi_N$, and the error dynamics forms a feedback loop only with the downstream part $\Xi_0$,
a block diagram of which is illustrated in the right of Fig.~\ref{fig:hier_e}.
In the latter case, the error dynamics has no direct interaction with $\Xi_0$.
However, there exists a free parameter $\hat{A}_0$ that satisfies $\hat{F}_0=0$ only when $\im{P_0}$ is $A$-invariant, which is a restrictive requirement.
Therefore, we hereinafter consider only the former case.

When~\eqref{eq:Eiz} is satisfied, the other parameters should be chosen so as to reduce the norm of the transfer matrix from $\xi_0,\xi_1,\ldots,\xi_N$ to $e$ given by
\[
 G_e:= (sI-\hat{A}_e)^{-1}[\hat{F}_0\ \hat{F}_1\ \cdots\ \hat{F}_N]
\]
while satisfying~\eqref{eq:paras_e}.
A reasonable policy to determine the parameters is to reduce the norm of the input matrices through
\begin{equation}\label{eq:choice}
\begin{array}{cl}
\hat{A}_0 \hs \in \displaystyle{\argmin_{X} \|\hat{F}_0(X)\|,}\\
(\hat{A}_i,\hat{R}_i) \hs \in \displaystyle{\argmin_{(X,Y)} \|\hat{F}_i(X,Y)\|,\quad i=1,\ldots,N}\\
\end{array}
\end{equation}
with an appropriate norm $\|\cdot\|$ where
\[
\hat{F}_0(X):=AP_0-P_0X,\quad \hat{F}_i(X,Y):= AP_i-P_iX-P_0Y.
\]
Note that this choice yields the exact hierarchical model decomposition with~\eqref{eq:repcon} in Theorem~\ref{thm:rep} if the conditions~\eqref{eq:excond1} and~\eqref{eq:excond2} in Theorem~\ref{thm:ex} hold.
In this sense, the system~\eqref{eq:hier_e} with the parameters~\eqref{eq:Eiz} and~\eqref{eq:choice} can be regarded as a generalization of hierarchical model decomposition.
Accordingly, we define \emph{robust hierarchical model decomposition.}
\begin{defin}[Robust Hierarchical Model Decomposition]
The system in~\eqref{eq:hier_e} with the parameters~\eqref{eq:paras_e},~\eqref{eq:Eiz}, and~\eqref{eq:choice} is said to be a robust hierarchical model decomposition of~\eqref{eq:cl_sys}.
\end{defin}

There always exists a robust hierarchical model decomposition for any system and choice of clusters as claimed by the following proposition.

\begin{prop}\label{prop:ex_rob_hier}
There always exists a robust hierarchical model decomposition of~\eqref{eq:cl_sys} for any cluster set.
\end{prop}

Based on the result, we subsequently discuss distributed design with a robust hierarchical model decomposition.

\subsection{Distributed Design for Indecomposable Systems}
Through the obtained robust hierarchical model decomposition, distributed design can be achieved by designing subcontrollers for each subsystem as long as the global subcontroller, which corresponds to the downstream part, can cope with the error signal.
This fact is described by the following theorem.
\begin{theorem}[Stabilization under Approximation Error]\label{thm:error}
Consider a robust hierarchical model decomposition~\eqref{eq:hier_e}.
Let $\hat{K}_1,\ldots,\hat{K}_N$ be controllers such that the closed-loop systems~\eqref{eq:ui} are internally stable.
Moreover, let $\hat{K}_0$ be a controller such that the closed-loop system
\[
 \left\{
 \begin{array}{cl}
 \dot{\xi}_0 \hs = \hat{A}_0\xi_0+\hat{E}_0e+B_0\hat{u}_0\\
 \dot{e} \hs = \hat{A}_ee + \hat{F}_0\xi_0\\
 \hat{u}_0 \hs = \hat{K}_0(C_0\xi_0+C_0P_0^{\sf T}e)
 \end{array}
 \right.
\]
is internally stable.
Then the controller composed of~\eqref{eq:ctrls} and functional observers~\eqref{eq:linob} stabilizes the clustered interconnected system~\eqref{eq:cl_sys}.
\end{theorem}

Theorem~\ref{thm:error} implies that the local subcontrollers can be designed without any concern about approximation error as long as the downstream part is stabilized by the global subcontroller, for design of which robust control~\cite{Zhou1996Robust} can apply.

Further, the following proposition shows that the functional observer~\eqref{eq:exlinob} still works for robust hierarchical model decomposition.
\begin{prop}\label{prop:ob_error}
Consider a robust hierarchical model decomposition~\eqref{eq:hier_e}.
Then the system~\eqref{eq:exlinob} is a functional observer of $C_i\xi_i$ for~\eqref{eq:hier_e} as well.
\end{prop}

The idea of the construction is almost the same as that in Proposition~\ref{thm:exlinob}.
The only difference is that $\phi_i$ is an estimation of $P_i^{\sf T}P_0\xi_0+P_i^{\sf T}e$ instead of $P_i^{\sf T}P_0\xi_0$.
Thus, we can estimate $C_i\xi_i$ using $\phi_i$ even when approximation errors are present.


\section{Numerical Examples}
\label{sec:num}


\subsection{Motivating Example Revisited}

Consider the motivating example.
As each local internal controller $\hat{K}_i$ in~\eqref{eq:ctrls} for $i=1,2,3$, we employ a linear quadratic regulator (LQR) under the state weight $Q_i=I_{r_i}\otimes\mathrm{D}(q_{\theta},q_{\omega})$ with $(q_{\theta},q_{\omega})=(1,10^4)$
and the input weight $R_i=10^2I_{r_i}$ with a state observer whose observer gain is determined according to the LQR method under the state weight $10^3I_{2r_i}$ and the input weight $R_i$.
Similarly, the global subcontroller $\hat{K}_0$ in~\eqref{eq:ctrls} is designed based on the LQR method under $Q_0=I_{N}\otimes\mathrm{D}(q_{\theta},q_{\omega})$ and $R_0=10^2I_N$ with the state observer.
For implementation of local subcontrollers, the functional observer~\eqref{eq:exlinob} is employed.

\begin{figure}[t]
\centering
\subfloat[][Free response.]{\includegraphics[width=.98\linewidth]{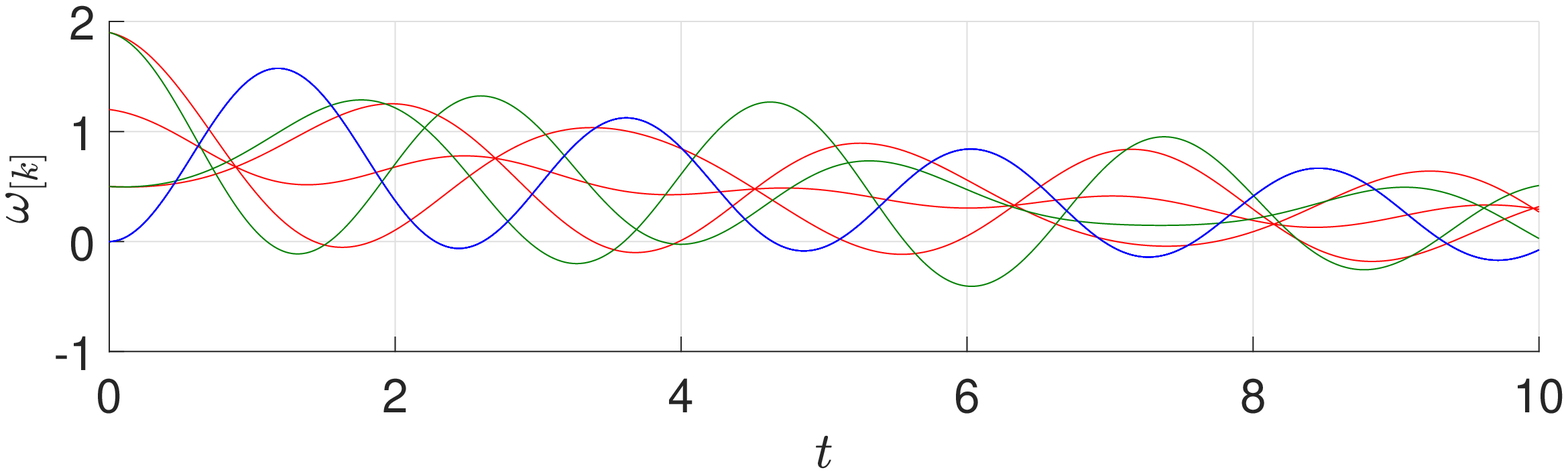}\label{subfig:free}}\newline
\subfloat[][Response only with the local subcontrollers.]{\includegraphics[width=.98\linewidth]{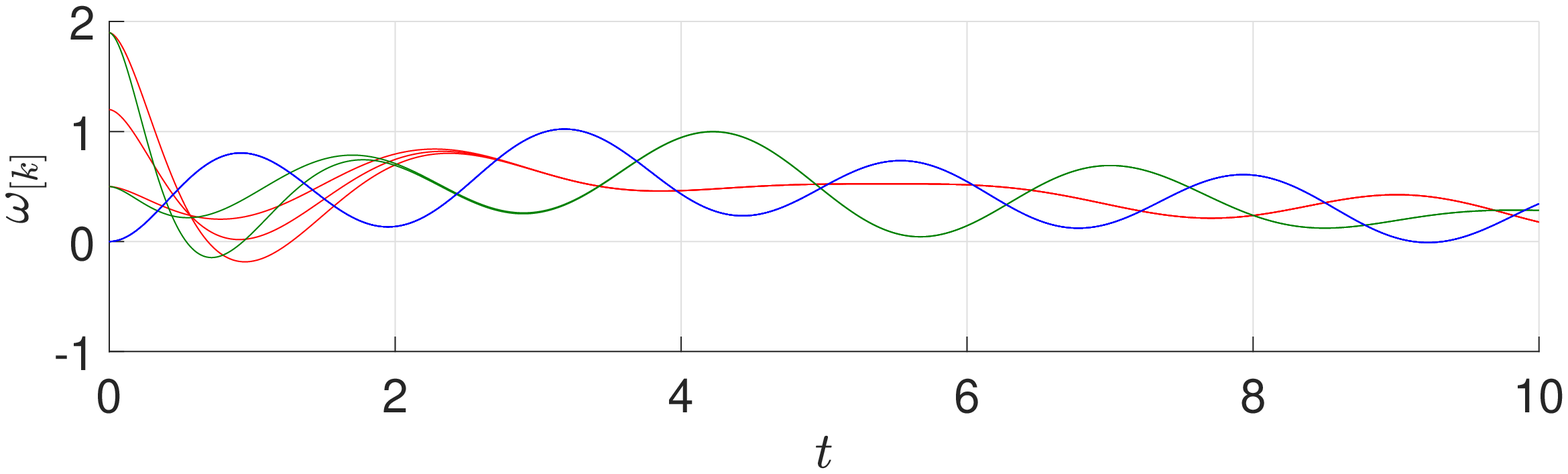}\label{subfig:local}}\newline
\subfloat[][Response only with the global subcontroller.]{\includegraphics[width=.98\linewidth]{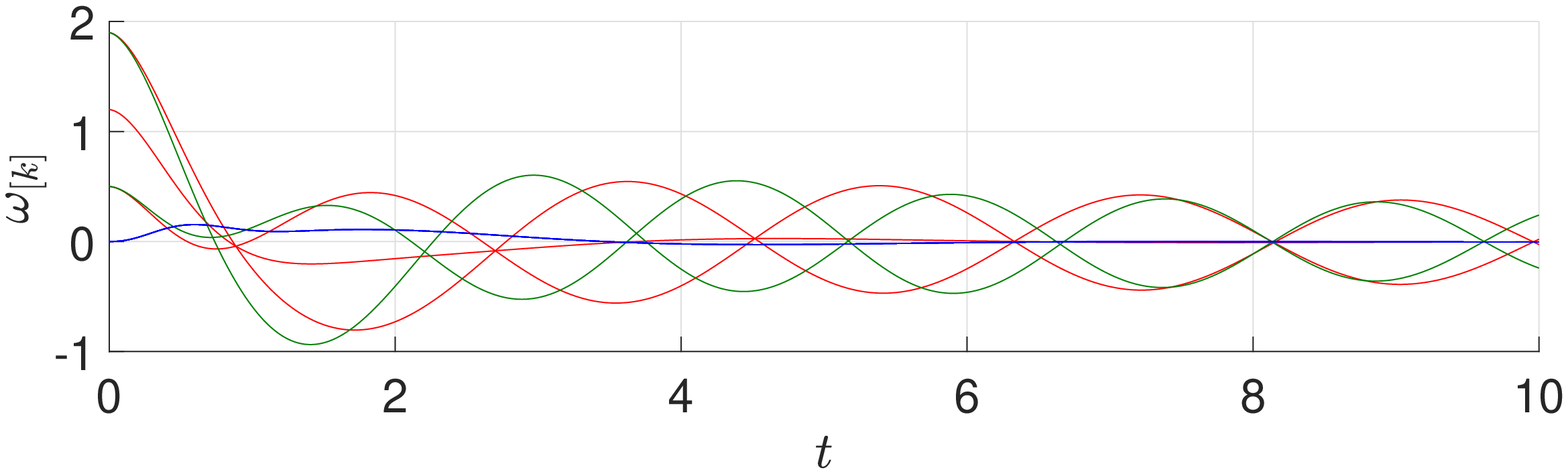}\label{subfig:global}}\newline
\subfloat[][Response with the glocal controller.]{\includegraphics[width=.98\linewidth]{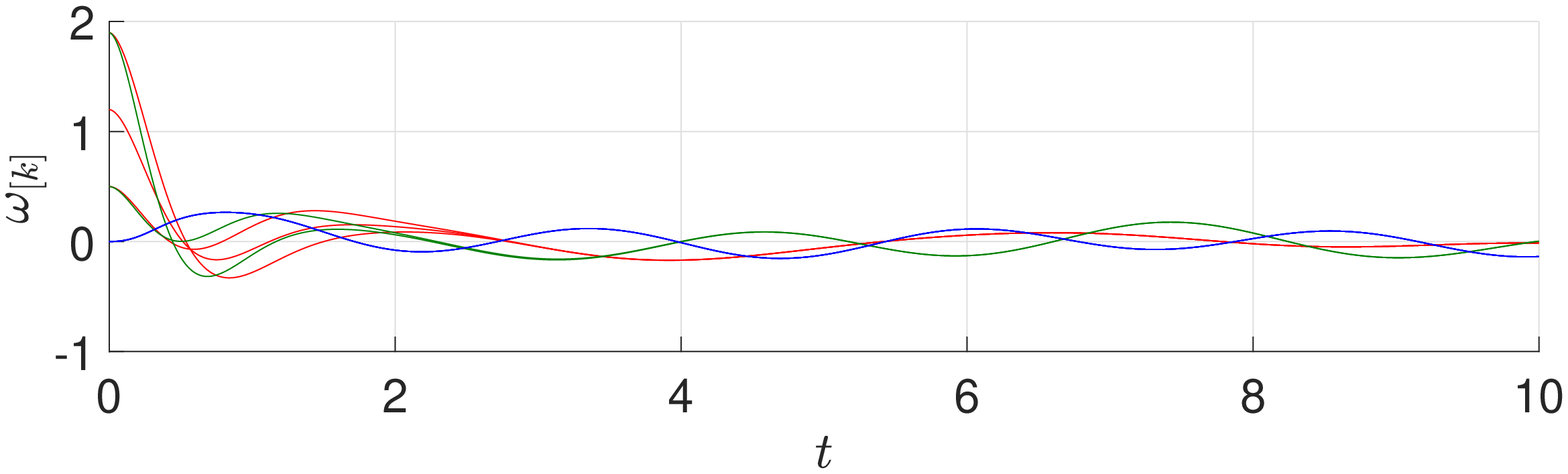}\label{subfig:glocal}}
\caption[]{Responses of the second-order network system with different control policies.}
\label{fig:graph_sec}
\end{figure}

The responses under the same initial condition as that of Fig.~\ref{fig:hier_sec_ex} are illustrated in Fig.~\ref{fig:graph_sec},
where Fig.~\ref{subfig:free}, Fig.~\ref{subfig:local}, Fig.~\ref{subfig:global}, and Fig.~\ref{subfig:glocal} correspond to the cases in which no controllers, only the local subcontrollers, only the global subcontroller, and the glocal controller is implemented, respectively.
It is observed in Fig.~\ref{subfig:local} that stationary interarea oscillation remains.
In Fig.~\ref{subfig:global}, local oscillation inside $\mathcal{I}_1$ and $\mathcal{I}_2$ cannot be suppressed only with the global subcontroller although interarea oscillation is removed.
In contrast to the above two cases, both of global/local behaviors can be suppressed by using the glocal controller as illustrated in Fig.~\ref{subfig:glocal}.
This result evidences the effectiveness of the glocal structure.

\begin{figure}[t]
\centering
\includegraphics[width=.95\linewidth]{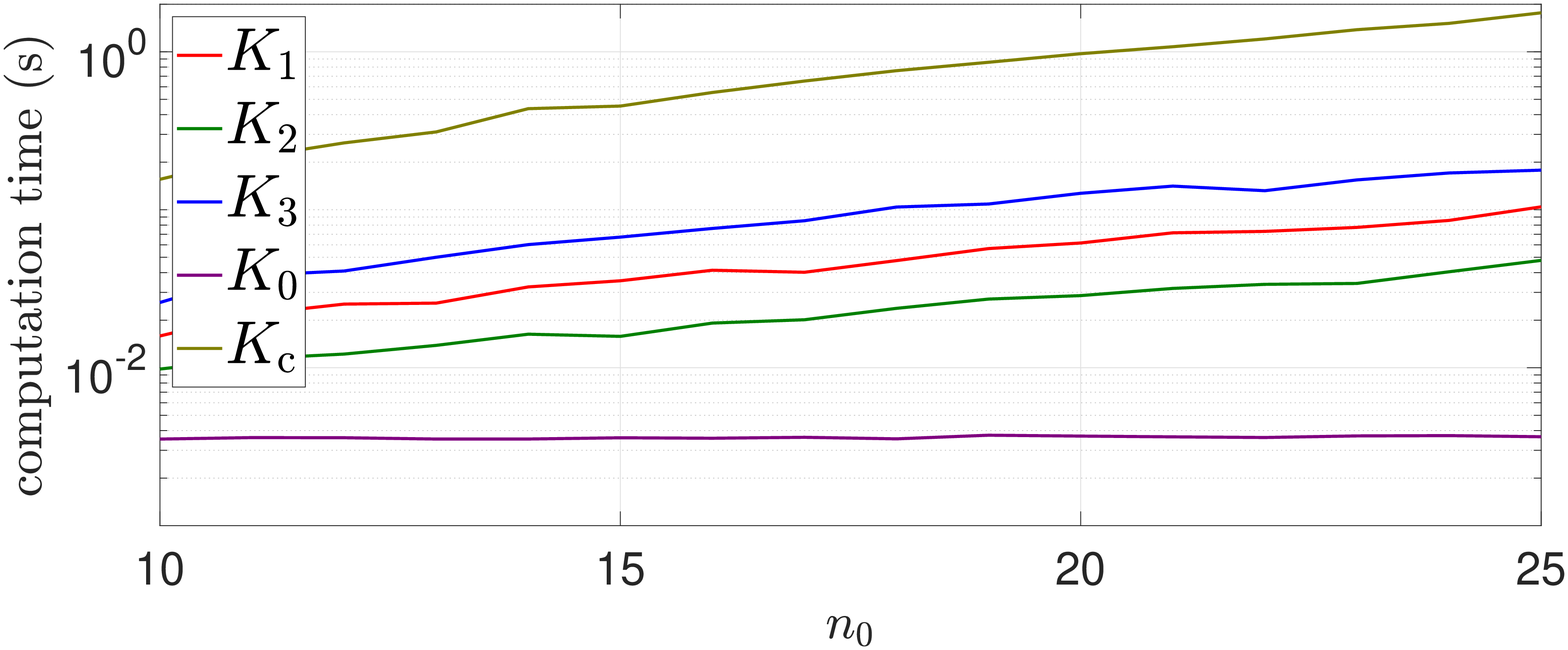}
\caption{Average computation times for designing the local subcontrollers $K_1,K_2,K_3$, the global subcontroller $K_0$, and the centralized controller $K_{\rm c}$ for varied $n_0$ ranging from $10$ to $25$ on a logarithmic scale.}
\label{fig:comp}
\end{figure}

Next, we confirm scalability of controller design by comparing computation times elapsed for designing a glocal controller and a centralized controller.
Consider increasing the number of subsystems inside each cluster in Fig.~\ref{fig:mot}.
Let $n_0$ be an index to determine the size of the system and set the number of components to $N_0=9n_0$.
Consider the three clusters constructed as
\begin{equation}\label{eq:clu_ex2}
\begin{array}{l}
 \mathcal{I}_1=\{1,\ldots,3n_0\},\quad \mathcal{I}_2=\{3n_0+1,\ldots,5n_0\},\\
 \mathcal{I}_3=\{5n_0+1,\ldots,9n_0\}
\end{array}
\end{equation}
and let the parameters of the components in each cluster be the same as those in Sec.~\ref{subsec:ex}.
We consider the centralized controller ${\rm col}(u_i)_{i=1}^N=K_{\rm c}{\rm col}(y_i)_{i=1}^N$ with a dense information structure.
For the centralized controller design, we employ the LQR method under the state weight $Q_{\rm c}=I_n\otimes\mathrm{D}(q_{\theta},q_{\omega})$ and the input weight $R_{\rm c}=10^2I_N$ with the state observer where $n$ is the dimension of all the states.
We also suppose that the global/local subcontrollers are designed in accordance with the previous ones.
The average computation times elapsed for designing the controllers with varied $n_0$ ranging from 10 to 25 are depicted in Fig.~\ref{fig:comp} on a logarithmic scale.
It can be observed that the computation time is significantly reduced through hierarchical model decomposition, which indicates scalability of the proposed distributed design.
Standard model reduction techniques, e.g., balanced truncation method~\cite{Antoulas2005Approximation} and singular perturbation method~\cite{Kokotovic1999Singular}, are ineffective in this case.
The Hankel singular values $\sigma_i$ of the transfer matrix from ${\rm col}(u_i)$ to ${\rm col}(y_i)$ for $n_0=20$ are given by
\[
\sigma_i \in \{1.3,1.6,1.7,2.4,2.5,+\infty\},\quad i=1,\ldots,360,
\]
where the unique singular value $+\infty$ corresponds to the semistable pole with the eigenvector $\mathds{1}_{N_0}\otimes [1\ 0]^{\sf T},$ which represents the direction of synchronized angles.
Because the ratio between the maximum singular value except for $+\infty$ and the minimum one is not very large,
all states of the entire system are irreducible through the balanced truncation method even if the semistable pole is disregarded.
Further, since the time scales of the global/local behaviors are not very different as seen in Fig.~\ref{subfig:free}, the singular perturbation approach is unprofitable for this system.
Indeed, using only the global subcontroller, which corresponds to the case where the local dynamics is reduced as a fast dynamics, cannot suppress the local oscillation as shown in Fig.~\ref{subfig:global}. 
Therefore, the computation time for designing $K_{\rm c}$ cannot efficiently be reduced using the traditional methods.

\begin{figure}[t]
\centering
\includegraphics[width=0.98\linewidth]{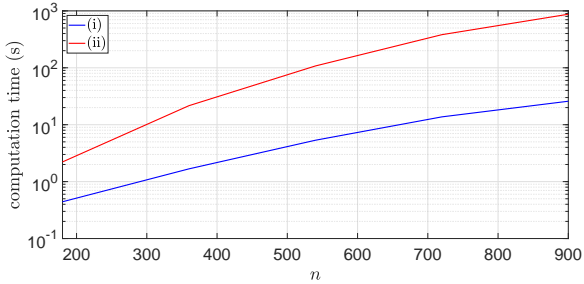}
\caption{Average computation times of the clustering algorithm for varied $n$, the dimension of the entire state space, ranging from $180$ to $900$ on a logarithmic scale.}
\label{fig:comp_al}
\end{figure}

We next investigate computational complexity of the proposed clustering algorithm.
Consider two situations:
\begin{enumerate}[(i)]
\item The size of the components increases under a fixed number of clusters.
\item The number of clusters increases under a fixed number of components in each cluster.
\end{enumerate}
Set $N_0=9n_0$.
In case (i), the parameters of the components in each cluster are set to be identical, which leads to the three clusters in~\eqref{eq:clu_ex2}.
In case (ii), the parameters of the components are set to give the clusters as
\[
 \begin{array}{l}
 \mathcal{I}_{3i+1}=\{9i+1,9i+2,9i+3\},\ \mathcal{I}_{3i+2}=\{9i+4,9i+5\},\\
 \mathcal{I}_{3i+3}=\{9i+6,9i+7,9i+8,9i+9\},\ i=0,\ldots,n_0-1.
 \end{array}
\]
The average computation times for the clustering algorithm are illustrated in Fig.~\ref{fig:comp_al} where the number of components $N_0$ ranges from $90$ to $450$ and the dimension of the entire state space $n$ ranges from $180$ to $900$.
It is observed that the computational complexity in case (i) is small compared with case (ii).
The computationally most expensive part in the algorithm is the calculation of the controllability matrix with respect to the pair $(A,P_i)$.
The overall computation time in case (ii), where the number of the clusters is $3n_0$, is longer than that in case (i), where the number of clusters is three.


\subsection{NPCC system}
\label{subsec:NPCC}

To illustrate the practical relevance of our proposed control structure, we consider the 48-machine NPCC system~\cite{Chow2013Power}, a model of the power grid in New York and the neighboring areas.
The NPCC 140-bus, 48-machine, 233-branch model can be found in the Power System Toolbox~\cite{Sauer2017Power}.
The interconnection parameters in~\eqref{eq:inter} are calculated based on its tie-line parameters.
Although each generator is modeled as the second-order system described by~\eqref{eq:second} as in the previous example,
the parameters are not strictly homogeneous.
Hence, there does not exist an exact hierarchical model decomposition for any non-trivial cluster set, and hence the proposed clustering algorithm cannot be applied to this system.
Accordingly, as a given cluster set, we employ the nine clusters depicted by~Fig.~3.5 in~\cite{Chow2013Power},
which is obtained through the coherency-based aggregation.
We apply the robust version of hierarchical model decomposition proposed in Sec.~\ref{sec:rob}.
Every internal subcontroller is designed as an LQR controller with a state observer.

\begin{figure}[t]
\centering
\subfloat[][Frequency deviation without any controller.]{\includegraphics[width=.48\linewidth]{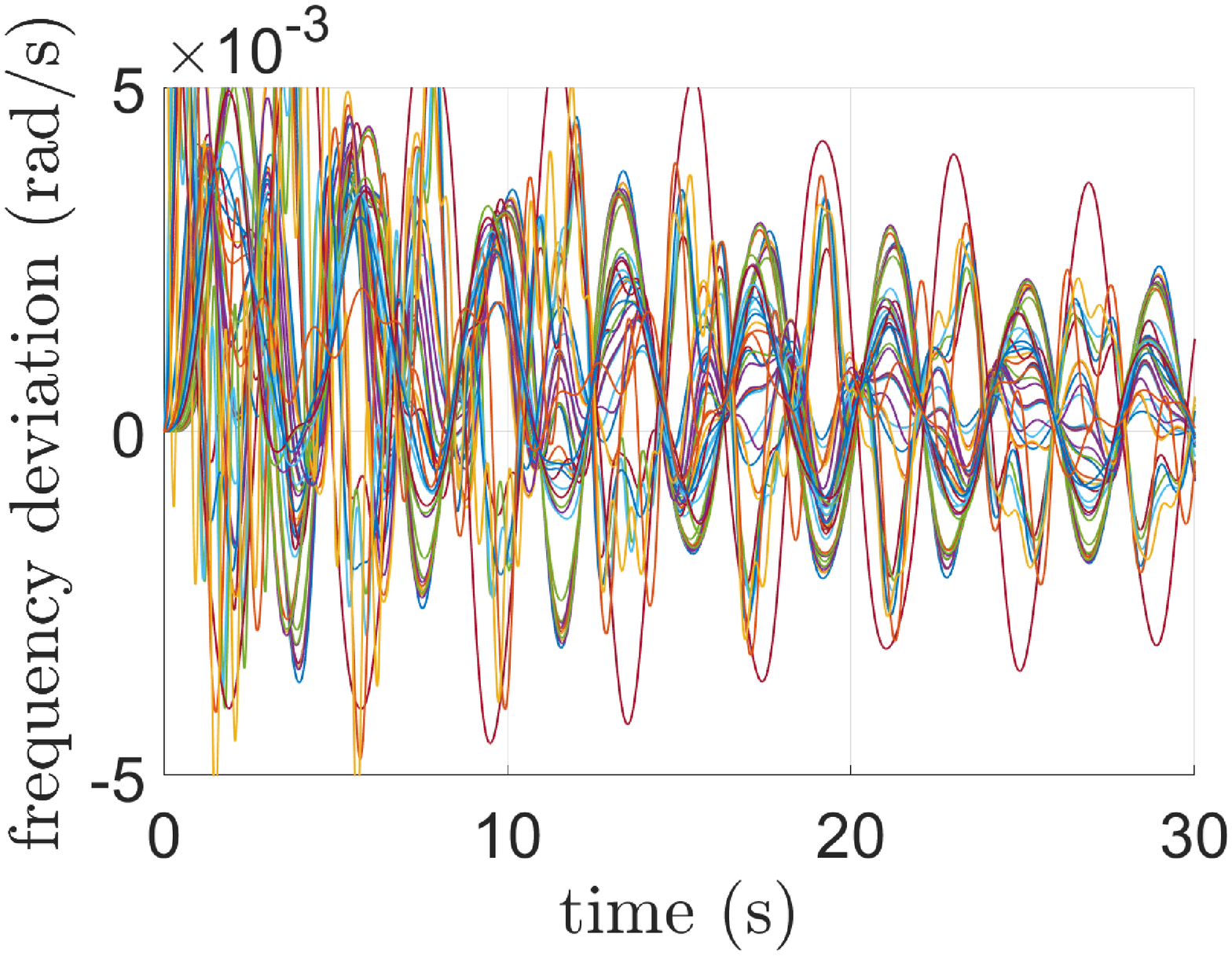}\label{subfig:free_NPCC}} \quad
\subfloat[][Frequency deviation only with the local subcontrollers]{\includegraphics[width=.48\linewidth]{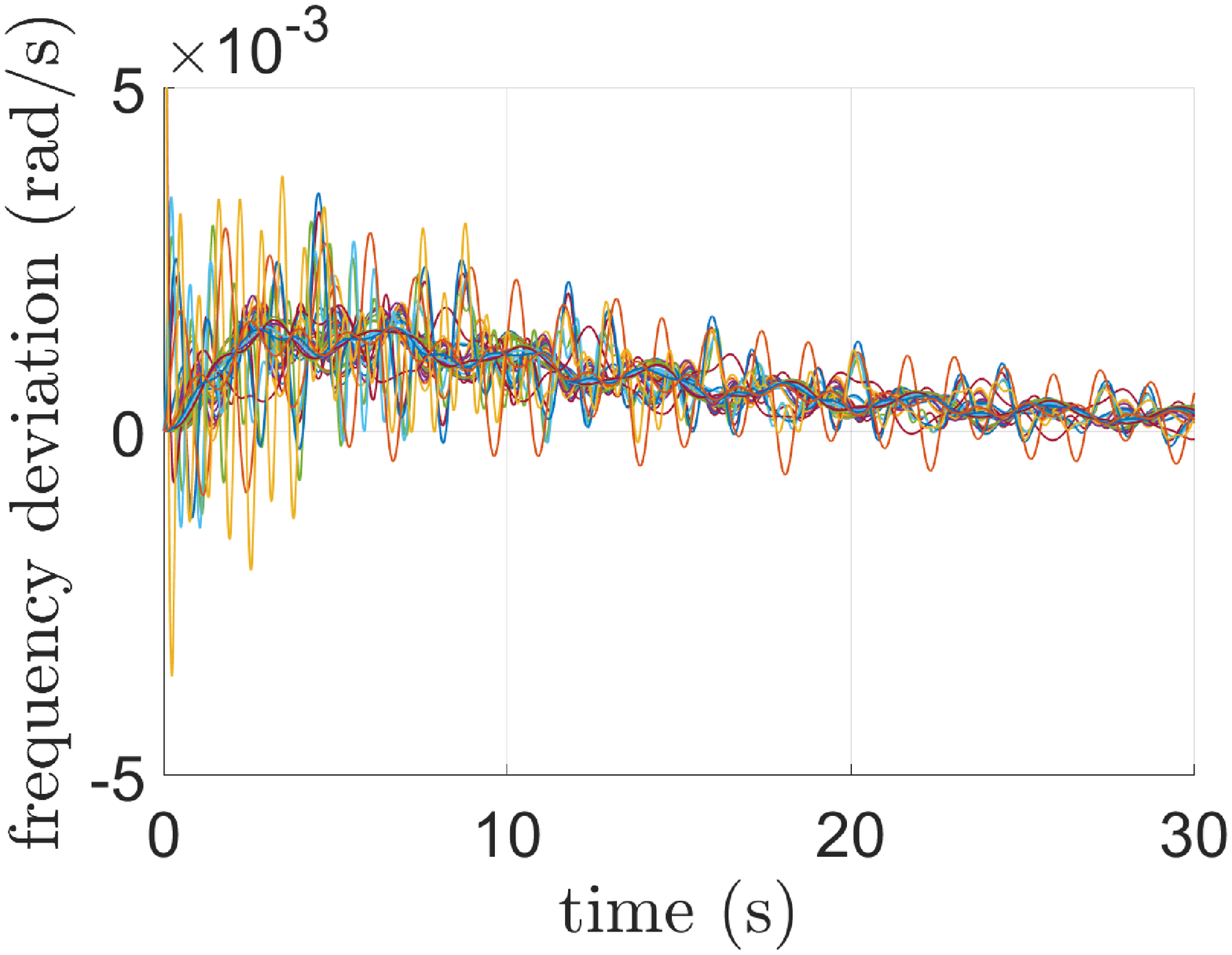}\label{subfig:local_NPCC}}\\
\subfloat[][Frequency deviation only with the global subcontroller.]{\includegraphics[width=.48\linewidth]{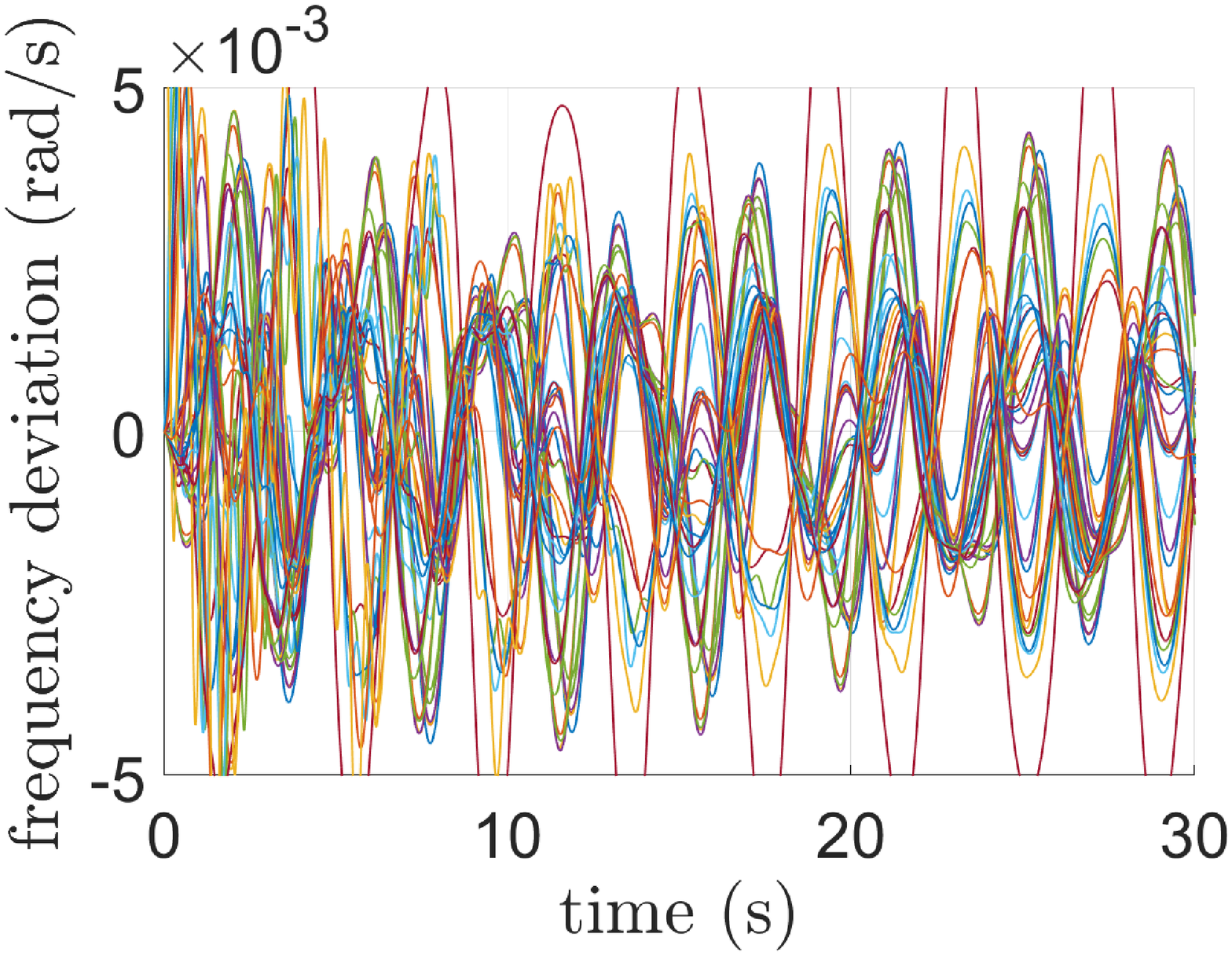}\label{subfig:global_NPCC}} \quad
\subfloat[][Frequency deviation with the glocal controller]{\includegraphics[width=.48\linewidth]{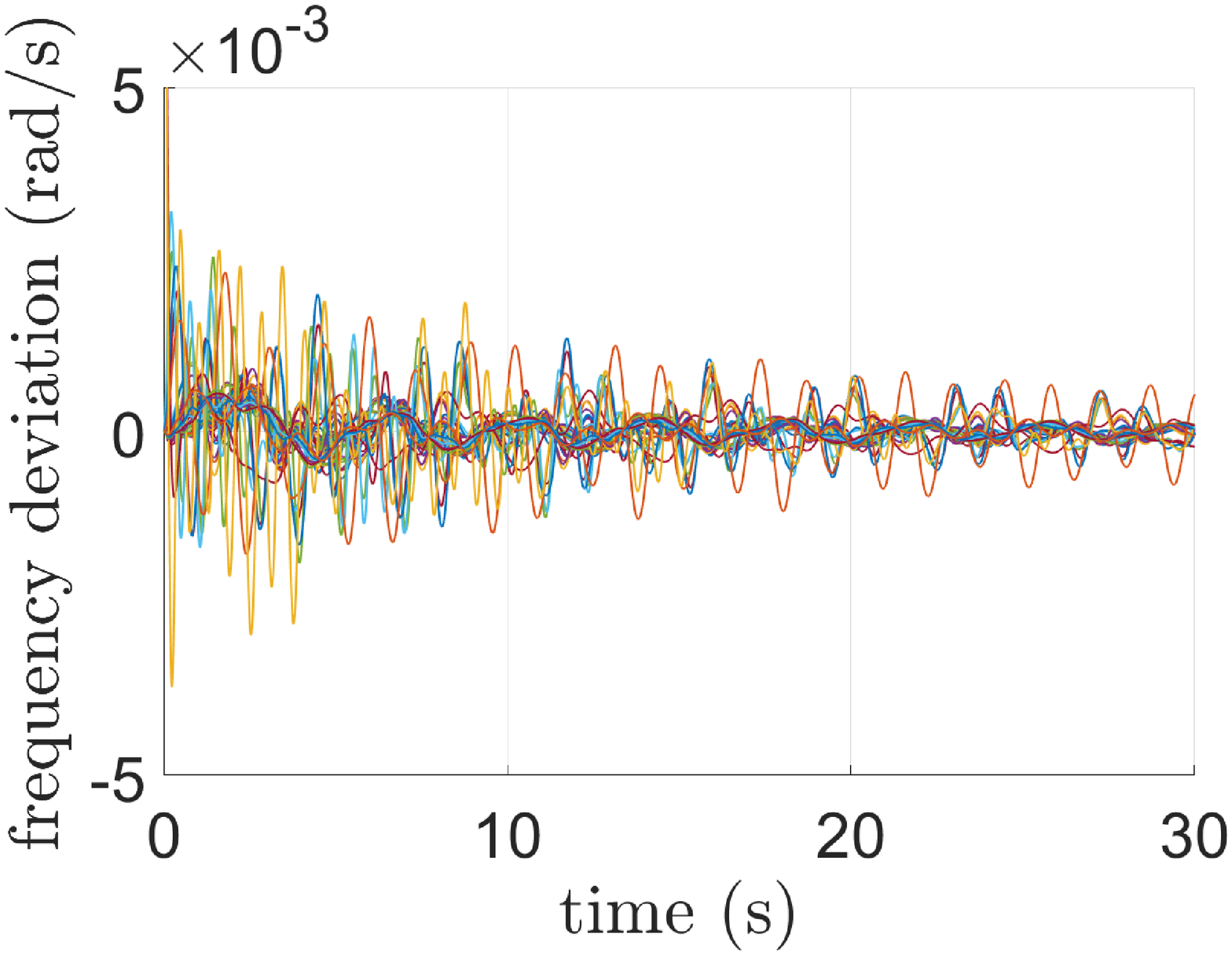}\label{subfig:glocal_NPCC}}
\caption[]{Responses of the NPCC testbed with different controllers.}
\label{fig:res_NPCC}
\end{figure}

\begin{figure}[t]
\centering
\includegraphics[width=0.98\linewidth]{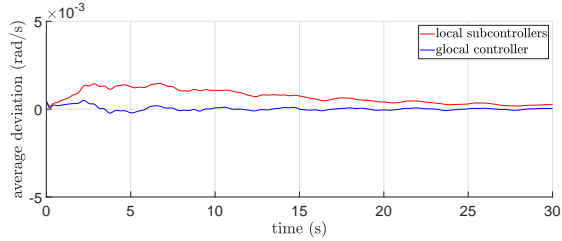}
\caption{Average frequency deviations of all generators with the local subcontrollers and with the glocal controller.}
\label{fig:average_fre}
\end{figure}

The frequency deviations of all generators are depicted in Fig.~\ref{fig:res_NPCC},
where Figs.~\ref{subfig:free_NPCC},~\ref{subfig:local_NPCC},~\ref{subfig:global_NPCC}, and~\ref{subfig:glocal_NPCC} show the responses without any controller, only with the local subcontrollers, only with the global subcontroller, and with the proposed glocal controller, respectively.
It can be observed that local oscillations are efficiently suppressed with local subcontrollers.
The average frequency deviations of all generators in the cases only with the local subcontrollers and with the glocal controller are depicted in Fig.~\ref{fig:average_fre}.
As shown in this figure, the excitation of the slow global dynamics, which remains in Fig.~\ref{subfig:local_NPCC}, is suppressed through the global subcontroller.
The result highlights the potential effectiveness of the proposed glocal control for practical systems.

\section{Conclusion}
\label{sec:conc}

In this paper, distributed design of glocal controllers has been proposed for large-scale network systems.
The proposed idea is to transform the original system into a cascade structured system, called hierarchical model decomposition.
Thanks to this structure, stability of the entire system can be guaranteed by designing subcontrollers each of which stabilizes the corresponding subsystem.
We have provided a condition for existence of the hierarchical model decomposition, a specific representation, a clustering method, and a robust extension.

Future research directions on the proposed framework include development of a clustering method that can handle approximation errors.
The proposed algorithm can result in a conservative decomposition for this case.
Moreover, although we have confined our attention only to the glocal structure,
distributed design of controllers with other particular information structures is an open problem.

\appendix[Proof of Theorems]

\section{Proofs of Theorems}
\begin{proof}
\emph{Proof of Theorem~\ref{thm:ex}:}
Sufficiency is shown in the proof of Theorem~\ref{thm:rep} by construction.
We here show necessity.
Assume that $\Xi$ is a hierarchical model decomposition.
Let $x_0 \in \im{P_0}$ and $x(0)=x_0, \xi_0(0)=P_0^{\dagger}x_0, \xi_i(0)=0$ for $i=1,\ldots,N$.
Then $P_0P_0^{\dagger}x_0 = x_0$ and $x(0) = P_0 \xi_0(0) + \sum_{i=1}^N P_i \xi_i(0)$.
Because $\Xi$ is a hierarchical model decomposition, $x(t)=P_0\xi_0(t)$ holds for any $t\geq 0$ with $\hat{u}_i=0$ for any $i$.
Thus $x(t) = e^{At}x_0 \in \im{P_0}$ for any $t\geq 0$.
Since $\im{P_0}$ is a closed subspace, $\lim_{t \rightarrow 0} (e^{At}x_0-x_0)/t = Ax_0 \in \im{P_0}$.
Because $x_0$ is arbitrary in $\im{P_0}$, $\im{P_0}$ is an invariant subspace of $A$.
Hence $\mathcal{R}(A,P_0) \subset \im{P_0}$.
Similarly, it can be shown that $\mathcal{R}(A,P_i) \subset \im{P_0}+\im{P_i}$ for $i=1,\ldots,N$ by taking $x(0) \in \im{P_i}$.
\end{proof}

\vspace{3mm}

\begin{proof}
\emph{Proof of Theorem~\ref{thm:rep}:}
We first show sufficiency.
As a preparation, we show that there exist matrices $\hat{A}_i$ and $\hat{R}_i$ such that~\eqref{eq:repcon} holds when the conditions~\eqref{eq:excond1} and \eqref{eq:excond2} are satisfied.
From~\eqref{eq:excond1}, $A\,\im{P_i} \subset \im{P_i}+\im{P_0}$
for $i=1,\ldots,N$.
Hence there exist $X_i$ and $X_{0i}$ such that $AP_i=P_iX_i+P_0X_{0i}.$
Moreover, from~\eqref{eq:excond2}, $\im{P_0}$ is $A$-invariant and hence there exists $X_0$ such that $AP_0=P_0X_0.$
Thus the condition~\eqref{eq:repcon} holds with $\hat{A}_i=X_i,\hat{R}_i=X_{0i}, \hat{A}_0=X_0$.

Let us assume that the condition~\eqref{eq:repcon} holds.
Define the error signal $\textstyle{e:=x-P_0\xi_0-\sum_{i=1}^NP_i\xi_i}$
and then
\[
 \begin{array}{cl}
 \dot{e} \hs = Ax-P_0(\hat{A}_0\xi_0+\sum_{i=1}^N\hat{R}_i\xi_i)-\sum_{i=1}^NP_i\hat{A}_i\xi_i\\
  \hs = Ax-P_0\hat{A}_0\xi_0-\sum_{i=1}^N(P_0\hat{R}_i+P_i\hat{A}_i)\xi_i\\
  \hs = Ax-AP_0\xi_0-\sum_{i=1}^NAP_i\xi_i =Ae.
 \end{array}
\]
When $x(0) = \sum_{i=1}^N P_i \xi_i(0) + P_0 \xi_0(0)$ holds, $e(0)=0$ and hence $e(t)=0$ for any $t\geq 0$ and $\hat{u}_i$.
Thus $\Xi$ is a hierarchical model decomposition.

We next show the necessity part.
Note that, from Theorem~\ref{thm:ex} $\mathcal{R}(A,P_0)\subset \im{P_0}$ holds.
We first show that $AP_0-P_0\hat{A}_0=0,$ which is equivalent to $\hat{A}_0=P_0^{\dagger}AP_0$ under $\mathcal{R}(A,P_0)\subset \im{P_0}$.
When $\hat{u}_i=0$ and $\xi_i(0)=0$, $\xi_i(t)=0$ for any $t\geq 0$.
Then $x(t) = P_0\xi_0(t)$
for any $t \geq 0$ for arbitrary $\hat{u}_0$ provided that $x(0)=P_0\xi_0(0)$.
Define $e_0:= P_0^{\dagger}x-\xi_0$ and then $\dot{e}_0=P_0^{\dagger}AP_0e_0 + (P_0^{\dagger}AP_0-\hat{A}_0)\xi_0.$
Since $e_0(t)=0$ for any $t \geq 0$, $\dot{e}_0(0)=0$ for any $\xi_0$.
Therefore the kernel of $P_0^{\dagger}AP_0-\hat{A}_0$ contains the entire space, which leads to $\hat{A}_0=P_0^{\dagger}AP_0$.
Similarly, it can be shown that $AP_i-P_0\hat{R}_i-P_i\hat{A}_i=0$ for $i=1,\ldots,N$.
\end{proof}

\vspace{3mm}

\begin{proof}
\emph{Proof of Theorem~\ref{thm:linob}:}
From the necessary and sufficient condition of functional observers~\cite[Lemma~2]{Fortmann1972Design}, $\mathbf{A}_i$ is stable and there exists a matrix $U_{i0}$ such that
\begin{equation}\label{eq:cond_linob}
\arraycolsep=2pt
\begin{array}{l}
 U_{i0} \left[
 \begin{array}{cc}
 \hat{A}_i & 0\\
 \hat{R}_i & \hat{A}_0
 \end{array}
 \right] -\mathbf{A}_iU_{i0}=\mathbf{E}_{i}C_i[I\ P_i^{\sf T}P_0],\\
 \left[\mathbf{B}_{i}\ \mathbf{D}_i\right] =
 U_{i0}\left[
 \begin{array}{cc}
 B_i & 0\\
 0 & B_0
 \end{array}
 \right],\ 
 [C_i\ 0]=\mathbf{C}_iU_{i0}+\mathbf{F}_iC_i[I\ P_i^{\sf T}P_0]
 \end{array}
\end{equation}
for any $i=1,\ldots,N$.
Let $U_{i0}=[U_i\ U_0]$ and define
\[
 \epsilon_i:= U_i\xi_i+U_0\xi_0-\phi_i.
\]
Then because $\Phi_i$ is a functional observer, the dynamics of $\epsilon_i$ can be represented as $\dot{\epsilon}_i = \mathbf{A}_i\epsilon_i$, which is stable.
Moreover, we have
\[
\begin{array}{cl}
 \psi_i \hs = \mathbf{C}_i\phi_i + \mathbf{F}_iC_i[I\ P_i^{\sf T}P_0][\xi_i^{\sf T}\ \xi_0^{\sf T}]^{\sf T}\\
  \hs = \mathbf{C}_i([U_i\ U_0][\xi_i^{\sf T}\ \xi_0^{\sf T}]^{\sf T} - \epsilon_i)+\mathbf{F}_iC_i[I\ P_i^{\sf T}P_0][\xi_i^{\sf T}\ \xi_0^{\sf T}]^{\sf T}\\
  \hs = C_i\xi_i-\mathbf{C_i}\epsilon_i
\end{array}
\]
in view of the third identity in~\eqref{eq:cond_linob}.
Thus the entire closed-loop system composed of $\Xi$ and $\{K_i\}_{i=0}^N$ can be described by
\begin{equation}\label{eq:hier_ob}
\left\{
\begin{array}{l}
 \left\{
 \begin{array}{cl}
 \dot{\epsilon}_i \hs = \mathbf{A}_i \epsilon_i\\
 \dot{\xi}_i \hs = \hat{A}_i\xi_i+B_i\hat{u}_i\\
 \hat{u}_i \hs = \hat{K}_i(C_i\xi_i-\mathbf{C}_i\epsilon_i)
 \end{array},\quad i=1,\ldots,N
 \right.\\
 \begin{array}{cl}
 \dot{\xi}_0 \hs =\hat{A}_0\xi_0+\sum_{i=1}^N\hat{R}_i\xi_i+B_0u_0\\
 \hat{u}_0 \hs =\hat{K}_0C_0(\xi_0+\sum_{i=1}^NP_0^{\sf T}P_i\xi_i).
 \end{array}
\end{array}
\right.
\end{equation}
From the cascade structure of~\eqref{eq:hier_ob} and the assumption on stability of every closed-loop system, the entire system is internally stable.
Because the original state can be represented by superposition of the states in~\eqref{eq:hier_ob}, the original system with the controllers is also internally stable.
\end{proof}

\vspace{3mm}

\begin{proof}
\emph{Proof of Proposition~\ref{thm:exlinob}:}
The whole observer composed of $\Phi_1,\ldots,\Phi_N$ can be represented by
\[
 \left\{
 \begin{array}{cl}
 \dot{\phi} \hs = \diag{A_i}\phi+\diag{A_i-\hat{A}_i}\hat{x}+\diag{L_i}v+P_0B_0\hat{u}_0\\
 \dot{\hat{x}} \hs = \diag{A_i}\hat{x}+\diag{B_i}{\rm col}(\hat{u}_i)+\diag{L_i}v+P_0B_0\hat{u}_0\\
 \psi \hs = -\diag{C_i}\phi+y
 \end{array}
 \right.
\]
with $\phi:= {\rm col}(\phi_i), \hat{x}:= {\rm col}(\hat{x}_i), \psi:={\rm col}(\psi_i)$ where the measurement signals are represented by
\[
 \left[
 \begin{array}{c}
 y\\
 v
 \end{array}
 \right]=\left[
 \begin{array}{cc}
 \diag{C_i} & \diag{C_i}P_0\\
 M & MP_0
 \end{array}
 \right]
 \left[
 \begin{array}{c}
 {\rm col}(\xi_i)_{i=1}^N\\
 \xi_0
 \end{array}
 \right].
\]
It suffices to show that the matrix
\[
 U:= \left[
 \begin{array}{cc}
 0 & P_0\\
 I & P_0
 \end{array}
 \right]
\]
satisfies the conditions
\begin{equation}\label{eq:firstcond}
\begin{array}{l}
 U\left[
 \begin{array}{cc}
 \diag{\hat{A}_i} & 0\\
 \left[\hat{R}_1\ \cdots \hat{R}_N\right] & \hat{A}_0
 \end{array}
 \right]-\left[
 \begin{array}{cc}
  \diag{\hat{A}_i} & \diag{A_i-\hat{A}_i}\\
 0 & \diag{A_i}
 \end{array}
 \right]U\vspace{1mm}\\
 = \left[
 \begin{array}{cc}
 0 & \diag{L_i}\\
 0 & \diag{L_i}
 \end{array}
 \right]\left[
 \begin{array}{cc}
 \diag{C_i} & \diag{C_i}P_0\\
 M & MP_0
 \end{array}
 \right],
\end{array}
\end{equation}
 and
\[
\begin{array}{l}
 \left[
 \begin{array}{cc}
 0 & P_0B_0\\
 \diag{B_i} & P_0B_0
 \end{array}
 \right]=
 U
 \left[
 \begin{array}{cc}
 \diag{B_i} & 0\\
 0 & B_0
 \end{array}
 \right], \vspace{1mm} \\
 
 [\diag{C_i}\ 0]= -[\diag{C_i}\ 0]U+[\diag{C_i}\ \diag{C_i}P_0].
\end{array}
\]
The second and third identities obviously hold.
Regarding the first condition, the right-hand side of~\eqref{eq:firstcond} is described by
\begin{equation}\label{eq:RHS}
\begin{array}{cl}
 {\rm (RHS)} \hs =
[I\ I]^{\sf T}
\diag{L_i}M
 \left[
 I\ P_0
 \right] \vspace{1mm} \\
 
 \hs =
 [I\ I]^{\sf T}
 (A-\diag{A_i})
 \left[
 I\ P_0
 \right]
\end{array}
\end{equation}
and the left-hand side is described by
\[
\arraycolsep=2pt
 \begin{array}{l}
 {\rm (LHS)} \\ = \left[
 \begin{array}{c}
 I\\
 I
 \end{array}
 \right]\left[P_0 \left[\hat{R}_1\ \cdots\ \hat{R}_N\right]-\diag{A_i-\hat{A}_i}\ P_0\hat{A}_0-\diag{A_i}P_0\right].
 \end{array}
\]
Since $\Xi$ is a hierarchical model decomposition, $P_0\hat{R}_i=AP_i-P_i\hat{A}_i$ for $i=1,\ldots,N$ and $P_0\hat{A}_0=AP_0$.
Therefore
\[
 {\rm (LHS)} =
 [I\ I]^{\sf T}
 \left[A-\diag{A_i}\ AP_0-\diag{A_i}P_0\right],
\]
which is equal to~\eqref{eq:RHS}.
\end{proof}

\vspace{3mm}

\begin{proof}
\emph{Proof of Proposition~\ref{lem:suf}:}
We show that $A\,\im{P_0^{(1)}} \subset \im{P_0}.$
Noting that $\mathcal{R}(A,[P_2\ \cdots\ P_N])\subset \im{P_0^{(1)}}+\im{[P_2\ \cdots\ P_N]}$
from~\eqref{eq:excond1},
we have
\[
\begin{array}{cl}
 A\,\im{P_0^{(1)}} \hs \subset A\,\im{P_0}\\
  \hs \subset A\,\mathcal{R}(A,[P_2\ \cdots\ P_N]) \\
  \hs \subset \mathcal{R}(A,[P_2\ \cdots\ P_N])\\
\hs \subset \im{P_0^{(1)}}+\im{[P_2\ \cdots\ P_N]}
\end{array}
\]
because of~\eqref{eq:sing} and $A$-invariance of the controllable subspace.
Moreover, since $\im{P_0^{(1)}} \subset \im{P_1}$, we have
\[
\begin{array}{cl}
 A\,\im{P_0^{(1)}} \hs \subset A\,\im{P_1}\\
 \hs \subset \mathcal{R}(A,P_1)\\
 \hs \subset \im{P_1} + \im{P_0}
\end{array}
\]
from~\eqref{eq:excond1}.
Therefore, it follows that
\[
 \begin{array}{cl}
 A\,\im{P_0^{(1)}} \hs \subset (\im{P_0^{(1)}}+\im{[P_2\ \cdots\ P_N]})\cap (\im{P_1} + \im{P_0})\\
 \hs= (\im{P_0^{(1)}}\cap\im{P_1})\oplus(\im{[P_2\ \cdots\ P_N]}\cap\im{P_0})\\
 \hs = \im{P_0}.
\end{array}
\]
Similarly, we can show the same inclusion property for the other clusters.
Hence, $\im{P_0}$ is $A$-invariant and the condition~\eqref{eq:excond2} holds.
\end{proof}

\vspace{3mm}

\begin{proof}
\emph{Proof of Theorem~\ref{thm:subopt}:}
Let $\{\mathcal{I}_i\}$ be the cluster set produced by Algorithm~1.
From Lemma~\ref{lem:clusters}, any cluster set in $\mathfrak{G}(\{\mathcal{I}^{(0)}_i\}_{i=1}^{N^{(0)}})$ can be generated from $\{\mathcal{I}_i\}$.
Since the number of clusters increases by partition, the claim holds.
\end{proof}

\vspace{3mm}

\begin{proof} 
\emph{Proof of Lemma~\ref{lem:clusters}:}
We prove the claim by induction.
It suffices to show $\mathfrak{G}(\{\mathcal{I}^{(\tau)}_i\}) \subset \mathfrak{G}(\{\mathcal{I}^{(\tau+1)}_i\})$ for any $\tau$.
Take a cluster set $\{\mathcal{I}_i\}_{i=1}^N$ that belongs to $\mathfrak{G}(\{\mathcal{I}^{(\tau)}_i\})$.
Because the clusters satisfy~\eqref{eq:excond1} and~\eqref{eq:excond2},
it suffices to show that $\{\mathcal{I}_i\}$ belongs to $\mathfrak{F}(\{\mathcal{I}^{(\tau+1)}_i\})$.
Let $k\in \{1,\ldots,N^{(\tau)}\}$ be the minimum index such that
\[
 \mathcal{R}(A,P_k^{(\tau)})\not\subset \im{P_k^{(\tau)}}+\im{P_0^{(\tau)}}
\]
where $P_k^{(\tau)}:=f_{P_i}(\mathcal{I}_k^{(\tau)})$ and $P_0^{(\tau)}:=f_{P_0}(\{\mathcal{I}_i^{(\tau)}\})$.
Because $\mathcal{I}_k^{(\tau)}$ is not partitioned at the $\tau$th step, there exists $j\in\{1,\ldots,N^{(\tau+1)}\}$ such that $\mathcal{I}_j^{(\tau+1)}=\mathcal{I}_k^{(\tau)}.$
Hence
\[
 \mathcal{R}(A,P_j^{(\tau+1)})=\mathcal{R}(A,P_k^{(\tau)}) \subset \im{P_k^{(\tau)}}+\im{P_0^{(\tau+1)}}
\]
since~\eqref{eq:excond1} holds for $j$.
Because $\{\mathcal{I}_i\}\subset \mathfrak{F}(\{\mathcal{I}_i^{\tau}\})$, there exists an index set $\mathcal{L}\subset \{1,\ldots,N\}$ such that $\cup_{l\in \mathcal{L}}\mathcal{I}_l=\mathcal{I}_k^{(\tau)}=\mathcal{I}_j^{(\tau+1)}$.
Because $\{\mathcal{I}_i\}$ satisfies~\eqref{eq:excond1}, we have
\[
 \mathcal{R}(A,P_\mathcal{L})=\mathcal{R}(A,P_k^{(\tau)}) \subset \im{P_k^{(\tau)}}+\im{P_0}
\]
where $P_\mathcal{L}$ is composed of $P_l:=f_{P_i}(\mathcal{I}_l)$ for $l\in \mathcal{L}$ and $P_0:=f_{P_0}(\{\mathcal{I}_i\})$.
By taking the intersection, we have
\[
 \mathcal{R}(A,P_k^{(\tau)}) \subset \im{P_k^{(\tau)}}+(\im{P_0^{(\tau+1)}} \cap \im{P_0}).
\]
Taking the projection onto the orthogonal subspace of $\im{P_k}^{(\tau)}$, we have
\[
 \pi_{\im{P_k}^{(\tau) \perp}}\mathcal{R}(A,P_k^{(\tau)}) \subset \pi_{\im{P_k}^{(\tau) \perp}}(\im{P_0^{(\tau+1)}} \cap \im{P_0}).
\]
The optimality of $\{\mathcal{I}_i^{(\tau+1)}\}$ in the problem~\eqref{eq:opt} yields
\[
 \pi_{\im{P_k}^{(\tau) \perp}}(\im{P_0^{(\tau+1)}} \cap \im{P_0}) = \pi_{\im{P_k}^{(\tau) \perp}}\im{P_0^{(\tau+1)}}.
\]
This identity and the relation $\cup_{l\in \mathcal{L}}\mathcal{I}_l=\mathcal{I}_k^{(\tau)}=\mathcal{I}_j^{(\tau+1)}$ imply that \{$\mathcal{I}_i\}$ is a partition of $\{\mathcal{I}_i^{(\tau+1)}\}$.
\end{proof}

\vspace{3mm}

\begin{proof}
\emph{Proof of Proposition~\ref{prop:invariance}:}
Let $P_i$ and $P_0$ be the corresponding matrices of the initial cluster set and $P_i'$ and $P_0'$ be the ones corresponding to the expanded cluster set.
Because $\im{P_0}\subset\im{P_0'}$, we have
\[
 \mathcal{R}(A,P_i) \subset \im{P_i}+\im{P_0}\subset \im{P_i}+\im{P_0'},\quad i\notin\mathcal{J}.
\]
Similarly, because $\im{P_\mathcal{J}} \subset \im{P_\mathcal{J'}},$
where $P_\mathcal{J}$ is composed of $P_j$ for $j\in \mathcal{J}$,
the condition~\eqref{eq:sing} holds for $i\notin \mathcal{J}$.
\end{proof}

\vspace{3mm}

\begin{proof}
\emph{Proof of Proposition~\ref{prop:ex_clustering}:}
From Assumption~\ref{assum:identical_clustering}, the condition of Assumption~\ref{assum:identical} holds.
Thus Theorem~\ref{thm:ex} leads to the claim.
\end{proof}

\vspace{3mm}
\begin{proof}
\emph{Proof of Proposition~\ref{prop:ex_rob_hier}:}
Because $\hat{F}_i$ for $i=0,1,\ldots,N$ are affine maps and any norm operator is convex, there exist $\hat{A}_0$ and $(\hat{A}_i,\hat{R}_i)$ that satisfy~\eqref{eq:choice} for any $i=0,1,\ldots,N$.
\end{proof}

\vspace{3mm}

\begin{proof}
\emph{Proof of Theorem~\ref{thm:error}:}
As in the proof of Theorem~\ref{thm:linob},
we can obtain an equivalent system
\[
\left\{
\begin{array}{l}
 \left\{
 \begin{array}{cl}
 \dot{\epsilon}_i \hs = \mathbf{A}_i \epsilon_i\\
 \dot{\xi}_i \hs = \hat{A}_i\xi_i+B_i\hat{u}_i\\
 \hat{u}_i \hs = \hat{K}_i(C_i\xi_i-\mathbf{C}_i\epsilon_i)
 \end{array},\quad i=1,\ldots,N
 \right.\\
 \begin{array}{cl}
 \dot{\xi}_0 \hs =\hat{A}_0\xi_0+\sum_{i=1}^N\hat{R}_i\xi_i+B_0u_0\\
 \dot{e} \hs = \hat{A}_ee + \hat{F}_0\xi_0\\
 \hat{u}_0 \hs =\hat{K}_0C_0(\xi_0+\sum_{i=1}^NP_0^{\sf T}P_i\xi_i+P_0^{\sf T}e).
 \end{array}
\end{array}
\right.
\]
From the cascade structure and the assumption on internal stability, the claim holds.
\end{proof}

\vspace{3mm}

\begin{proof}
\emph{Proof of Proposition~\ref{prop:ob_error}:}
By simple algebra, it can be confirmed that the matrix
\[
 U_e:=\left[
 \begin{array}{ccc}
 0 & P_0 & I\\
 I & P_0 & I
 \end{array}
 \right]
\]
satisfies the conditions for the functional observers of ${\rm col}(C_i\xi_i)$.
\end{proof}

\bibliography{sshrrefs}
\bibliographystyle{IEEEtran}

\begin{IEEEbiography}{Hampei Sasahara}(M'??)
Biography text here.
\end{IEEEbiography}

\begin{IEEEbiography}{Takayuki Ishizaki}(M'??)
Biography text here.
\end{IEEEbiography}

\begin{IEEEbiography}{Jun-ichi Imura}(M'??)
Biography text here.
\end{IEEEbiography}

\begin{IEEEbiography}{Henrik Sandberg}(M'??)
Biography text here.
\end{IEEEbiography}

\begin{IEEEbiography}{Karl Henrik Johansson}(M'??)
Biography text here.
\end{IEEEbiography}

\end{document}